\documentclass[aps,twocolumn, pra]{revtex4}

\usepackage{amsmath}
\usepackage{graphicx}

\usepackage{epsfig}
\usepackage{bm}

\newcommand{\be}{\begin{equation}}
\newcommand{\ee}{\end{equation}}

\newcommand{\beq}{\begin{eqnarray}}
\newcommand{\eeq}{\end{eqnarray}}

\def\H1{\widehat{H}_1}

\newcommand{\HH}{{\cal H}}

\begin{document}

\author{C. De Grandi$^1$}
\author{V. Gritsev$^2$}
\author{A. Polkovnikov$^1$}
\title{Quench dynamics near a quantum critical point: application to the sine-Gordon model}
\affiliation{$^1$Department of Physics, Boston University, 590 Commonwealth Avenue, Boston, MA 02215, USA\\
$^2$Department of Physics, University of Fribourg, Chemin du Mus\'{e}e 3, 1700 Fribourg, Switzerland}

\begin{abstract}
We discuss the quench dynamics near a quantum critical point focusing on the sine-Gordon model as a primary example.
We suggest a unified approach to sudden and slow quenches, where the tuning parameter $\lambda(t)$ changes in time as $\lambda(t)\sim \upsilon t^r$, based on the adiabatic expansion of the excitation probability in powers of $\upsilon$. We show that the universal scaling of the excitation probability can be understood through the singularity of the generalized adiabatic susceptibility $\chi_{2r+2}(\lambda)$, which for sudden quenches ($r=0$) reduces to the fidelity susceptibility. In turn this class of susceptibilities is expressed through the moments of the connected correlation
function of the quench operator. We analyze the excitations created after a sudden quench of the cosine potential using a combined approach of form-factors expansion and conformal perturbation theory for the low-energy and high-energy sector respectively. We find  the general scaling laws for the probability of exciting the system, the density of excited quasiparticles, the entropy and the heat generated after the quench. In the two limits where the sine-Gordon model maps to hard core bosons and free massive fermions we provide the exact solutions for the quench
dynamics and discuss the finite temperature generalizations.

\end{abstract}
\maketitle

\section{Introduction}

The Sine-Gordon (SG) model is one of the few examples of exactly
solvable models both in the classical and in the quantum case. In the
classical limit, an arbitrary initial perturbation splits into solitons
and breathers, which are infinitely long lived excitations (see e.g.
Ref.~\cite{novikov}). In the quantum regime these solitons and
breathers are quantized and form a discrete spectrum (for each
momentum state)~\cite{Zamolodchikov95, Mussardo-book}. This model
found numerous applications in various fields of physics including
statistical physics, condensed-matter physics, atomic physics and
high-energy physics. Its integrability allows to solve
various equilibrium and non-equilibrium situations and gain
important insights into phenomena ranging from classical and quantum
phase transitions~\cite{chaikin-lubensky, giamarchi, Sachdev_book}
to quench dynamics of coupled superfluids~\cite{cazalilla_quench,
gritsev:quench}.

For the purposes of this work we will keep in mind the applications of
this model to  one-dimensional quantum systems. Thus a natural
low-energy theory describing a one-dimensional interacting gas of
bosons or fermions is the Luttinger liquid model, which is nothing
but a scalar free $1+1$ dimensional bosonic theory~\cite{giamarchi}.
The Luttinger liquid is often unstable to various perturbations
opening a gap in the system coming either from the external potential
(periodic~\cite{buchler} or disordered~\cite{GS}) or the tunneling
coupling between multiple Luttinger liquids~\cite{giamarchi_ho}. In
these situations the most relevant perturbations appear in the form
of the lowest harmonic of the cosine potential allowed by the
symmetry~\cite{haldanelong, Sachdev_book} leading to the sine-Gordon
Hamiltonian:
\be\label{sgham0}
H=\frac{1}{2}\int dx
\left[(\partial_{x}\phi(x))^{2}+(\Pi(x))^{2}-4\lambda\cos(\beta\phi(x))\right].
\ee
In  cold-atom systems this Hamiltonian can be implemented through
optical lattices~\cite{buchler, nostro}, it also emerges naturally
in the context of interacting quantum wires
\cite{giamarchi},\cite{meyer_wigner}, mesoscopic superconducting
junctions \cite{gil_pair_junct}, spin
chains~\cite{giamarchi, Sachdev_book} and many others.

The recent experimental developments in cold-atoms \cite{Bloch2008_rmp,
experiments1} prompted a rapidly growing theoretical interest to
study the \textit{dynamics} of quantum systems.
A particular attention was paid to such issues as sudden quench
dynamics~\cite{cardy0quenches, cardy_temp, cazalillaLuttinger,
sengupta_sachdev,kollath_altman_quench,roux,faribault},
thermalization in integrable and nonintegrable
systems~\cite{olshanii_nature,rigolPRL,rigolPRLrelaxation,kollathThermalization}
and slow dynamics in isolated systems near quantum critical
points~\cite{toliapoint, zurekzoller, dziarmaga}. The latter works
illustrated the universal scaling of the density of quasiparticles
generated due to nonadiabatic transitions during the crossing of
quantum phase transitions. The universal properties of quantum critical
points in equilibrium extend to the dynamics and allow one to make
universal predictions on the defects and energy generation similar to
the Kibble-Zurek relations~\cite{kz1, kz2} for slow dynamics near
quantum critical
points~\cite{levitov,uma_quenchXY,sengupta2D,canevaIsing,uma_gapless,lviola,sengupta_nonlinearPRL,
sengupta_nonlinearPRB,umaMulticritical,balazs,rossini}.

In the preceding work~\cite{quench_short} we showed that a similar
universality also persists for a sudden quench dynamics near quantum
critical points if the quench amplitude is sufficiently small. There
we analyzed the general scaling behavior of different quantities
following a sudden quench of the amplitude $\lambda_f$ starting from
the critical point $\lambda_i=0$. We presented some general arguments for
the scaling of various quantities such as the probability of
exciting the system $P_{\rm ex}$, the density of excited
quasiparticles $n_{\rm ex}$, the (diagonal) entropy density
$S_d$~\cite{diag_entropy}, and the heat density $Q$~\cite{ap_heat},
with the quench amplitude and the system size. We showed that the
corresponding scaling laws are universal and can be described by
two critical exponents characterizing the quantum critical point:
the dynamical exponent $z$ and the correlation length exponent
$\nu$. We also argued that the scaling behavior for sudden quenches
is smoothly connected to the scaling behavior of similar quantities
for slow quenches, where the tuning parameter is turned on as an arbitrary
power of time~\cite{krishnendu, optimal_passage}: $\lambda(t)\sim
\upsilon t^r$, where $\upsilon$ is a small parameter. In
particular, in the limit $r\to 0$ this parameter becomes the quench
amplitude: $\upsilon=\lambda_f$.

The goal of this paper is to verify how the general arguments of
Ref.~\cite{quench_short} work in practice. We will use the
sine-Gordon model as a primary example. Performing explicit
calculations based on the adiabatic perturbation theory, form-factor
approach, and conformal perturbation theory we will
illustrate how these scalings are realized for the case of quenching the cosine
potential and highlight various subtleties which can emerge due to
possible ultraviolet or infrared divergencies. In the two limits
where the SG model reduces to free massive bosons and fermions we
will extend the analysis to higher dimensions and again verify the
general predictions of Ref.~\cite{quench_short}. In these two limits
we will be able to generalize the results to finite temperatures and
see how the quasiparticle statistics affects the scaling of various
quantities.

The paper is organized as follows: in Sec.~\ref{sec:gen} we review
the general scaling results presented in Ref.~\cite{quench_short}
introducing a generalized framework that allows us to describe
sudden quenches and slow quenches on the same footing. In
Sec.~\ref{sec:SGintro} we describe the basic properties of the SG
model. Then in Secs.~\ref{sec:FF} and \ref{sec:cpt} we analyze the
quench dynamics in the SG model using the adiabatic perturbation
theory. In Sec.~\ref{sec:exact} we present the exact analysis of
quenches in the two noninteracting limits of the SG model
corresponding to the free massive bosons and free massive fermions.
In Sec.~\ref{sec:temp} we furthermore extend this analysis to finite
temperature quenches. Finally  in Sec.~\ref{sec:geom_tens} we
discuss some interesting relations between quench dynamics, fidelity
susceptibility and geometric tensors.

\section{Generalized framework for the analysis of quench dynamics: hierarchy of susceptibilities }
\label{sec:gen}

In this section we will illustrate a generalized approach to study
the scaling laws of the quantities of interest after performing a
quench. As in Ref.~\cite{quench_short} we consider a $d$-dimensional
system described by a Hamiltonian $H(\lambda)=H_0+\lambda V$, where
$H_0$ is the Hamiltonian corresponding to a quantum critical point
(QCP) and $V$ is a relevant (or marginal) perturbation. In
particular, we consider the processes where the coupling changes
near the critical point  as a power law in time:
\beq\label{lambda_t}
\lambda(t)=\upsilon {\frac{t^r}{r!}}\Theta(t),
\eeq
where $\upsilon$ is a small parameter that we define as $\upsilon \equiv \frac{d^r \lambda}{d t^r}\mid_{t=0}$ and $\Theta$ is the step
function, so that the coupling starts changing with time as $t^r$
near the critical point. On the same footing we can consider the
opposite processes, where quenches stop at the critical point as:
\be
\lambda(t)=\upsilon {(t_f-t)^r\over r!}\Theta(t_f-t).
\ee
The analysis also extends to cyclic processes where the dynamical
evolution starts and ends at the quantum critical point. E.g.
$\lambda(t)\approx \upsilon {t^r(t_f-t)^r/ (r!\, t_f^r)}$. In all
these cases $\upsilon$ plays the role of the adiabatic parameter
such that $\upsilon\to 0$ corresponds to the adiabatic limit. In
particular, if the system is initially prepared in the ground state
then in the limit $\upsilon\to 0$ it will remain in the ground
state. The physical meaning of $\upsilon$ is quite transparent. For
sudden quenches $r=0$, it corresponds to the quench amplitude
$\upsilon=\lambda_f$; for linear quenches $r=1$, it describes the
quench rate $\upsilon=\dot\lambda$; for quadratic quenches $r=2$, it
describes the acceleration near the quantum critical point:
$\upsilon=\ddot\lambda$ and so. The smallness of $\upsilon$ suggests
the possibility of using it as an expansion parameter. Relying on
the adiabatic perturbation theory one can indeed write an expansion
in the powers of $\upsilon$. The details of this approach can be
found in Refs.~\cite{kolkata, ortiz_2008}. We  only quote the final
expression for the transition amplitude to the excited state
$|n\rangle$ as a result of this process (see e.g. Eq.~(19) in
Ref.~\cite{kolkata} or Eq. ~(18) in Ref.~\cite{ortiz_2008}):
\begin{widetext}
\beq\label{alpha00}
 &&\alpha_n(t_f)\approx \left[i {\langle n|\partial_t|0\rangle\over E_n(t)-E_0(t)}-{1\over E_n(t)-E_0(t)}
 {d\over dt}{\langle n|\partial_t|0\rangle\over E_n(t)-E_0(t)}+\dots\right]\mathrm e^{i(\Theta_n(t)-\Theta_0(t))}\Biggr|_{t_i}^{t_f}\nonumber\\
 &&=\left[i \dot\lambda{\langle n|\partial_\lambda| 0\rangle
 \over E_n(\lambda)-E_0(\lambda)}-\ddot\lambda{\langle n|\partial_\lambda|0\rangle\over
 (E_n(\lambda)-E_0(\lambda))^2}-\dot\lambda^2{1\over E_n(\lambda)-E_0(\lambda)}{d\over d\lambda}
 {\langle n|\partial_\lambda|0\rangle\over E_n(\lambda)-E_0(\lambda)}+\dots\right]\mathrm
 e^{i(\Theta_n(\lambda)-\Theta_0(\lambda))}\Biggr|_{\lambda_i}^{\lambda_f},\phantom{XX}
 \label{expansion1}
 \eeq
\end{widetext}
where $\Theta_n(t)=\int_0^t E_n(\tau)d\tau$ is the dynamical phase
(in general one should also add the Berry phase term) and $E_n$ is the
eigenenergy of the state $|n\rangle$ . We can anticipate that the
critical point will dominate the dynamics so that only one of the limits
$\lambda=\lambda_i$ (if $\lambda_i=0$) or $\lambda=\lambda_f$ (if
$\lambda_f=0$) should dominate the transition amplitude. From this
equation it is clear that the lowest non-vanishing time derivative
of $\lambda(t)$ at the critical point dominates the asymptotics of
the transition probability:
\be\label{alpha0}
|\alpha_n|^2\approx \upsilon^2 {|\langle n|\partial_\lambda|0\rangle|^2\over (E_n(0)-E_0(0))^{2r}}=
\upsilon^2 {|\langle n|V|0\rangle|^2\over (E_n(0)-E_0(0))^{2r+2}},
\ee
where all the matrix elements are evaluated at the quantum critical
point. Therefore we  find that the probability of excitations,
defined with respect to the instantaneous ground state, scales as:
\be
P_{\rm ex}=\sum_{n\neq 0}|\alpha_n|^2\approx L^d \upsilon^2 \chi_{2r+2}(0),
\label{pex_ups}
\ee
where
\be
\chi_{m}(\lambda)={1\over L^d}\sum_{n\neq 0}{|\langle n| V|0\rangle|^2\over (E_n(\lambda)-E_0(\lambda))^m}
\label{chim}
\ee
is the generalized adiabatic susceptibility of  order $m$. For $m=2$
($r=0$) $\chi_2=\chi_f$, where $\chi_f$ is the fidelity
susceptibility:
\be
\label{chif}
\chi_f={1\over L^d}\sum_{n\neq 0} |\langle 0|\partial_\lambda|n\rangle|^2=
{1\over L^d}\sum_{n\neq 0} { |\langle 0|V|n\rangle|^2\over (E_n(\lambda)-E_0(\lambda))^2}.
\ee
Thus $\chi_f$ describes the system's response to a sudden quench
(see Refs.\cite{quench_short} and \cite{gu_long}). For $m=4$ ($r=1$)
the linear adiabatic susceptibility $\chi_4$ describes the response
to a linear quench and so on. Let us note that the scaling dimension
of the susceptibilities $\chi_m(\lambda)$ is related to  the scaling dimension of the fidelity susceptibility via:
$\dim[\chi_m(\lambda)]=\dim[\chi_f(\lambda)]-z(m-2)=d-2/\nu-z(m-2)$.
If $\dim[\chi_m(\lambda)]$ becomes negative ($d\nu<2+z\nu(m-2)$)
then the susceptibility $\chi_m(\lambda)$ diverges at the quantum
critical point. Since $\dim[\lambda]=1/\nu$ the asymptotical
behavior of $\chi_m(\lambda)$ at small $\lambda$ is:
\be
\chi_m(\lambda)\sim {1\over |\lambda|^{2-d\nu+z\nu(m-2)}}.
\label{chi_mlam}
\ee
In finite size systems this divergence is cutoff by the system size
$L$ when the correlation length $\xi(\lambda)\sim 1/|\lambda|^\nu$
becomes comparable to $L$. Thus exactly at the quantum critical
point the generalized adiabatic susceptibility scales as
\be
\chi_m(0)\sim L^{2/\nu+z(m-2)-d}.
\label{chi_ml}
\ee
For $m=2$ this scaling agrees with that of the fidelity
susceptibility~\cite{venuti, alet}.  If the scaling dimension of
$\chi_m$ is positive then this generically indicates that this
susceptibility has a cusp singularity, meaning that the
asymptotics~(\ref{chi_mlam}) and (\ref{chi_ml}) become subleading.

Combining Eqs.~(\ref{pex_ups}) and (\ref{chi_ml}) we find that if
the condition $d\nu\leq 2(1+z\nu r)$ is fulfilled, then $P_{\rm ex}$
has the following asymptotic behavior in the adiabatic limit:
\be\label{Pexr}
P_{\rm ex}\approx \upsilon^2 L^{2/\nu+2 z r}.
\ee
Since the probability $P_{\rm ex}$ must always be smaller than
one, this scaling must be valid only for quenches of small amplitude
$|\upsilon|\ll 1/L^{1/\nu+zr}$. The physical meaning of this
requirement is that the correlation length  of the system associated
with this velocity $\xi(\upsilon)\sim 1/|\upsilon|^{\frac{\nu}{1+z r
\nu}}$ must be big compared to the system size $\xi\gg L$.
Physically the length scale $\xi(\upsilon)$ characterizes the
crossover in the response of the system from the sudden to the
adiabatic regimes similar to the Kibble-Zurek mechanism~\cite{kz1,
kz2}. I.e. for excitations with characteristic size (e.g.
wavelength) shorter than $\xi(\upsilon)$ the quenching process looks
adiabatic, while for excitations with size larger than
$\xi(\upsilon)$ it looks sudden. For linear quenches
$\xi(\upsilon)\sim 1/|\upsilon|^{\frac{\nu}{1+z \nu}}$ indeed
reproduces the characteristic Kibble-Zurek length scale~\cite{kz1,
kz2}.

Let us point that the crossover to the regime with $P_{\rm
ex}(\upsilon)\approx 1$ occurs precisely when the characteristic
length scale $\xi(\upsilon)$ becomes of the order of the system
size. This can be explained using a simple symmetry argument, which
is easier to understand for sudden quenches $r=0$. In this case the
quench corresponds to  projecting the ground state in the
critical state on the new basis. The quantity $1-P_{\rm ex}$ defines
the probability that the system remains in the ground state after
the quench. But if $\xi(\upsilon)<L$  the new ground
state has a well defined symmetry absent at the critical point. Thus
the probability to remain in the ground state should be close to zero.
Similar considerations apply to slow quenches,  $r>0$.
Thus for quenches of higher rate $|\upsilon|\gg 1/L^{1/\nu+zr}$ (but
still slow compared to the other microscopic scales of the system)
the probability of exciting the system is close to one and is no
longer a good quantity to characterize the system response (like the
overlap of two ground states becomes very close to zero and thus not
very informative). We argued \cite{quench_short} that in this case
the physically relevant and meaningful quantity to look at is the
density of generated excitations $n_{\rm ex}$ (quasiparticles if we
are dealing with nearly integrable models). Indeed the physical
reason why the global probability of excitation rapidly approaches
unity is that exciting even a single quasiparticle makes the new
state orthogonal to the initial state. Yet physically we anticipate
that exciting a single quasiparticle in the system should not change
drastically the properties of the system. In particular, it should
not affect the probability to excite the next quasiparticle. Since
we are typically dealing with few body operators by performing a
quench we couple only to the  many-body states characterized by few
quasiparticles. Thus  it was argued that the density of generated
quasiparticles should scale as:
\be\label{nex03}
n_{\rm ex}\sim \upsilon^2 L^{2/\nu-d+2 z r}
\ee
for $|\upsilon|\ll 1/L^{1/\nu+zr}$ and
\be
n_{\rm ex}\sim |\upsilon|^{d\nu/(z\nu r+1)}
\label{nex3}
\ee
in the opposite limit. Note that the second scaling  has a much
wider domain of applicability than the first one since there is no
requirement on vanishing quench amplitude with the system size. In
particular, for $r=1$  the result  (\ref{nex3}) agrees with the
Kibble-Zurek scaling predicted earlier in Refs.~\cite{toliapoint,
zurekzoller}. Also for nonlinear quenches $r>1$ the scaling
(\ref{nex3}) agrees with the results of Refs.~\cite{krishnendu,
optimal_passage}, noting that $\upsilon=(1/\tau)^r$ in the notations of
Ref.~\cite{krishnendu}. Let us also point out that, as  we will
illustrate here for the specific case of the sine-Gordon model
(Sec.~\ref{sec:cpt}), for $d\nu>2$ the nonanalytic asymptotic with
$|\upsilon|$ or $L$ in the scaling for $n_{\rm ex}$ and $P_{\rm ex}$
survives but becomes subleading. The leading asymptotic is then
analytic and it comes from the excitations of non-universal high
energy states.

The density of excitations is a very simple and intuitive object,
however it is not always well defined. Unless we deal with long
lived quasiparticles or other e.g. topological excitations, in
nonintegrable systems $n_{\rm ex}$ is not conserved in time  after
the quench because of various relaxation processes. For the same
reason $n_{\rm ex}$ is not a readily observable quantity. Hence one
needs to characterize the response of the system by other means. The
two natural quantities, which can be defined for any Hamiltonian
system, are the (diagonal) entropy~\cite{diag_entropy} and the heat
(or the non-adiabatic part of the energy, which is also the excess
energy above the new ground state of the quenched
Hamiltonian~\cite{ap_heat}). Because both quantities are extensive,
it is convenient to deal with their densities:
\beq
S_d&=&-{1\over L^d} \sum_{n} |\alpha_n|^2\log|\alpha_n|^2,\\
Q&=&{1\over L^d}\sum_n (E_n-E_0)|\alpha_n|^2=\sum_{n\neq 0} E_n |\alpha_n|^2. \label{q_def}
\eeq
Here the index $n$ enumerates the many-body eigenstates of the
Hamiltonian in the final state and $|\alpha_n|^2$, as written above in
Eqs. (\ref{alpha00}) and (\ref{alpha0})), is the probability of the transition to the $n$-state due
to the quench. The
scaling of the entropy is very similar (up to possible logarithmic
corrections) to the one of $P_{\rm ex}$ and $n_{\rm ex}$. The
advantage of the entropy over $P_{\rm ex}$ is that it is well
behaved and meaningful for large amplitude quenches $|\upsilon|\gg
1/L^{1/\nu+zr}$, where $P_{\rm ex}\to 1$. The scaling of the heat is
different from that of $n_{\rm ex}$ because of the term $E_n-E_0$ in
Eq.~(\ref{q_def}). If the quench ends at the critical point then one
finds that the scaling of the heat is described by $\chi_{2r+1}$:
$Q\sim \upsilon^2 \chi_{2r+1}$ so that
\be\label{Qex03}
Q\sim \upsilon^2 L^{2/\nu-d+z (2r-1)}
\ee
for $|\upsilon|\ll 1/L^{1/\nu+zr}$ and
\be
Q\sim |\upsilon|^{(d+z)\nu/(z\nu r+1)}
\label{Qex3}
\ee
for $|\upsilon|\gg 1/L^{1/\nu+zr}$.

Let us note that the susceptibilities $\chi_m$ are directly related
to the connected parts of the correlation functions
$G_\lambda(\tau)=\langle V(\lambda,\tau)V(\lambda,0)\rangle-\langle
V(,\lambda,0)\rangle^2$ by a simple generalization of the expressions
obtained in Ref.~\cite{venuti}:
\be
\chi_m(\lambda)={1\over L^d (m-1)!}\int_0^\infty \tau^{m-1}G_\lambda(\tau) d\tau.
\ee
For local perturbations $V=\sum_x V(x)$ it is obvious that if
$G(\tau)$ decays sufficiently fast at large $\tau$ then $\chi_m$ is
a non-negative number well defined in the thermodynamic limit
$L\to\infty$. For the sine-Gordon model we wrote the explicit
expressions of $\chi_1$ and $\chi_2$ in Eqs.~(\ref{chi_e}) and
(\ref{chi_f}). Using the Lehmann's representation of the correlation
function we can define the generating functions for the susceptibilities
$\chi_m(\lambda)$:
\beq
I(\alpha)&=&{1\over L^d}\int_0^\infty G_\lambda(\tau)\mathrm e^{-\alpha\tau} d\tau\nonumber\\
&=&{1\over L^d} \sum_{n\neq 0} {|\langle n|V|0\rangle|^2\over E_n(\lambda)-E_0(\lambda)+\alpha},
\eeq
such that
\be
I(\alpha)=\sum_m (-\alpha)^m\chi_m(\lambda).
\ee
This characteristic function clearly contains the information about the
scalings laws of the quantities discussed in this work for all
possible power law quenches.

Let us say a few words about the scaling in the regime where the
relevant adiabatic susceptibility has only a cusp singularity. In
this case the leading asymptotics of $\chi_m(\lambda)$ at
$\lambda\to 0$ is given by a nonuniversal constant, which in turn
defines the leading response of the system to the quench. Thus one
can expect the  scaling for the quantities $n_{\rm ex}$,
$S_d$, $Q$ to be analytic in $\upsilon$. Instead of Eq.~(\ref{nex03}) we
anticipate:
\be\label{nex04}
n_{\rm ex}\sim \upsilon^2 \chi_{2r+2}(0).
\ee
Similarly for the heat (in the case of a  quench ending at the QCP)
we expect $Q\sim \upsilon^2 \chi_{2r+1}(0)$. As we will show below
the non-analytic behavior of these quantities with $\upsilon$
survives, but becomes subleading and thus hardly identifiable either
numerically or experimentally.

It is interesting to note that these scaling results extend to the
situation of gapless systems even in the absence of the critical
point. Thus if we are dealing with gapless bosonic theory then one
simply needs to take the limit $\nu\to\infty$ in the expressions
above. It is clear from Eqs.~(\ref{chi_mlam}) and (\ref{chi_ml})
that for $m>2$ the susceptibility $\chi_m$ can still diverge in low
enough dimensions if  $d<z(m-2)$. This divergence is
actually the origin of the strong non-adiabaticity in
low-dimensional gapless systems~\cite{tolianature} (see also
Ref.~\cite{kollar}). Thus the existence of the quantum criticality
is not really needed to have a non-analytic scaling of various
observables with $\upsilon$. Following Ref.~\cite{kollar} let us
make the following observation. In the limit $\nu\to\infty$ in low
dimensions  the scaling of the  three quantities $n_{\rm ex}$,
$S_d$ and $Q$ is only sensitive to the quench time but not to the
actual quenching protocol. For instance for $d<z(2 r -1)$ the scaling of the heat is $Q\sim
|\upsilon|^{d+z\over zr}\sim 1/\tau^{d+z\over z}$, where $\tau\sim
(\lambda_f/\upsilon)^{1/r}$ is the quench time required to change
$\lambda$ from the initial value $\lambda_i=0$ to some fixed final value $\lambda_f$.
Thus the scaling of $Q$ with $\tau$ is not-sensitive to the power
$r$ with which the tuning parameter is quenched. On the other hand
for $d>z(2 r -1)$  the situation is opposite. The scaling of the heat
$Q\sim \upsilon^2\sim 1/\tau^{2r}$ is very sensitive to the power
with which the coupling $\lambda$ is turned on (off).
This suggests  that the optimum quench protocol, which gives the smallest excitation
level in the system for a given quench time $\tau$, corresponds to the
crossover  power $r=\frac{1}{2}(\frac{d}{z}+1)$. Higher powers of $r$ are to be avoided
because of the growth of the prefactor with $r$ \cite{optimal_passage}.
This argument extends the predictions of Ref.~\cite{optimal_passage} to generic gapless systems.

\section{The sine-Gordon model}
\label{sec:SGintro}

The SG-model is described by the Hamiltonian:
\be\label{sgham}
H=\frac{1}{2}\int dx
\left[(\partial_{x}\phi(x))^{2}+(\Pi(x))^{2}-4\lambda \cos(\beta\phi(x))\right],
\ee
where $\phi(x)$ and  $\Pi(x)$ are conjugated fields.
 Since this model is exactly solvable a great
amount of facts is known about it. It is conventional to introduce
the parameter $\xi$ ~\footnote{By doing this we avoid the
difficulties related to the differences in notations which are
common in the literature.}
\be\label{xi}
\xi=\frac{\beta^{2}}{8\pi-\beta^{2}}.
\ee
The spectrum of the SG Hamiltonian (\ref{sgham}) depends on the
value of $\beta$ (or $\xi$). We will be primarily interested in the
regime of $0<\beta^2<8\pi$ since in this range for a finite
$\lambda$ the system is in a gapped phase, while for $\lambda=0$
it corresponds to the gapless Luttinger liquid. Therefore the system
 in this regime undergoes a quantum
phase transition at  $\lambda=0$.
Furthermore for $4\pi<\beta^{2}<8\pi$
$(1<\xi<\infty)$, so called repulsive regime, the spectrum of the SG
Hamiltonian consists of solitons and antisolitons. At fixed small
$\lambda$ and $\beta^{2}=8\pi$ the systems undergoes a
Kosterlitz-Thouless transition to the Luttinger liquid regime, where
the cosine term is irrelevant. The point $\beta^{2}=4\pi$ ($\xi=1$),
known as Toulouse point,  maps to the free massive fermionic theory;
at this point solitons and antisolitons correspond to particles and
holes. For $0<\beta^{2}<4\pi$ $(0<\xi<1)$, so called attractive
regime, the spectrum in addition to the solitons and antisolitons
contains their bound states called breathers. The number of
different types of breathers depends on the interaction parameter
$\xi$ and is equal to the integer part of $1/\xi$. We denote a
breather by $B_{n}$ with $n=1,2, \ldots, [1/\xi]$. In the small
$\beta$ (or $\xi$) limit the SG model is well described by the
gaussian approximation, where one expands the $\cos(\beta\phi)$ term
in the Hamiltonian~(\ref{sgham}) to the quadratic order in $\phi$.
In this limit there is only one massive excitation, corresponding to
the lowest breather $B_{1}$. Solitons (kinks), antisolitons
(antikinks), and breathers are massive particle-like excitations.
The soliton and antisoliton  mass in terms of the parameters of the
Hamiltonian~(\ref{sgham}) was computed in
Ref.~[\onlinecite{Zamolodchikov95}]:
\beq\label{masss}
M_s=\left(\frac{\pi\Gamma\left(\frac{1}{1+\xi}\right)\lambda
}{\Gamma\left(\frac{\xi}{1+\xi}\right)}\right)^{(1+\xi)/2}
\frac{2\Gamma\left(\frac{\xi}{2}\right)}{\sqrt{\pi}\,
\Gamma\left(\frac{1+\xi}{2}\right)}.
\eeq
Breathers mass is related to the soliton mass via:
\beq\label{massb}
M_{B_{n}}=2M_s\sin\left(\frac{\pi \xi n}{2}\right).
\eeq
Note that for weak interactions (small $\xi$) the lowest breather
masses are approximately equidistant, $M_{B_n}\propto n$, which
suggests that these masses correspond to eigen-energies of a
harmonic theory. There is a direct analogy between the breathers in
the sine-Gordon model and the energy levels of a simple Josephson
junction~\cite{gritsev:quench}. In the Josephson junction (which is
described by the Hamiltonian~(\ref{sgham}) without the first
gradient term) the energy levels also become approximately
equidistant if the interaction (charging) energy is small. Let us
note that because the SG theory is explicitly relativistic invariant
the dynamical exponent $z$ is equal to 1. The soliton mass $M_s$
identifies the characteristic energy scale of the system, therefore
we expect for it a universal behavior around the critical point:
$M_s\sim |\lambda|^{z\nu}$, that defines the critical exponent $\nu$
for this theory (from Eq.~(\ref{masss}))
\beq\label{exp_nu}
 \nu=\frac{1+\xi}{2}
\eeq
or $\nu=4\pi/(8\pi-\beta^2)$,  which can be continuously  varied between $1/2$ and $\infty$.

We will focus on sudden quenches of the parameter $\lambda$,
$\lambda(t)=\lambda_f \Theta(t)$ and we will illustrate how the
general scalings laws for the density of excited quasiparticles
(Eqs. (\ref{nex03}) and (\ref{nex3})) can be obtained for this
model. As we discussed in the previous section and in
Ref.~\cite{quench_short} these scaling  laws are closely related to
the scaling of the fidelity susceptibility $\chi_f$. This
susceptibility can be expressed as an integral of the correlation
functions. For the SG model the latter can be accessed from  two
limits: the infrared limit (IR) describing long distance properties,
and the ultraviolet limit (UV) corresponding to short distances. The
approach based on the form factors is well suited to analyze the
former IR limit. To deal with the UV limit one can rely on the
conformal perturbation theory, which is based on the fact that at
short distances the SG model (as well as many other IR-massive
models) is free and conformal (it effectively reduces to the
Luttinger Liquid theory). These two approaches are complementary and
can be used depending on the physics of interest. As we will show
below for $ d\nu<2$ (or $\nu <2$ since $d=1$) the former form
factors approach reproduce the correct scalings of $P_{\rm ex}$ and
$n_{\rm ex}$, while for $\nu>2$ one should use the conformal
perturbation theory. Similar story is valid for heat but the IR and
UV dominated domains are defined according to whether
$(d+z)\nu=2\nu$ is bigger or smaller than $2$. In general, it is
rather difficult to relate the results from the two formalisms: in
order to get the UV physics from the form factors expansions one has
to sum up the large (if not infinite) number of contributions (see
Ref.~\cite{Mussardo-book} and the following Sec.~\ref{sec:FF} and
\ref{sec:cpt}). The converse is also true.  We will analyze both the
UV and IR contributions to the excitation probability and show that
their combination indeed reproduces the correct scaling asymptotics.

\section{Analysis of the low energy excitations based on the form factor expansion}
\label{sec:FF}

Let us start from the perturbative expression for the probability of
exciting the system based on the adiabatic perturbation theory.
Within the latter the transition amplitude to the state $|n\rangle$
is found as~\cite{kolkata}
\be
\alpha_n\approx-\int_0^{\lambda_f} d\lambda \langle n|\partial_{\lambda}|0\rangle.
\label{alpha_pert}
\ee
If the matrix element $\langle n|\partial_{\lambda}|0\rangle$ is not
singular at the QCP then this expression reduces to the result of
the first order of the conventional perturbation theory. The
expression (\ref{alpha_pert}) has an advantage that it symmetrically
treats both initial and the final values of the coupling $\lambda$
and gives convergent expressions even when the conventional
perturbation theory breaks down. From Eq.~(\ref{alpha_pert}) we find
\be
P_{\rm ex}(\lambda_f)\approx \sum_{n \neq 0}\int_0^{\lambda_f}\int_0^{\lambda_f} d\lambda_1 d\lambda_2
\langle 0|\partial_{\lambda_1}|n\rangle \langle n|\partial_{\lambda_2}|0\rangle.
\label{p_ex}
\ee
In general one has to sum over all intermediate states $|n\rangle$.
However, as we argued in \cite{quench_short}, if $d\nu<2$ (i.e.
$\xi<3$ in the SG model)  then $P_{\rm ex}$ is dominated by the low
energy excitations. Since solitons and breathers are massive, i.e.
gapped for $|\lambda|>0$, we can expect that the dominant
contribution to Eq.~(\ref{p_ex}) comes from single pairs of the
lowest energy excitations (solitons and antisolitons for $\xi>1$ and
$B_1$ breathers for $\xi<1$) with opposite momenta. Therefore the
probability of exciting the system and the number of excited
quasiparticles will become identical (up to a factor of two
reflecting that one excited state creates two solitons or
breathers). In this case  we can identify the sum over the states
$n$ as a sum over momenta $k$ and use:
\beq\label{alpha}
\alpha_k(\lambda_f)&=&-\int\limits_0^{\lambda_f} d\lambda' \langle
 k|\partial_{\lambda'}|0\rangle\\
 &=&2\int\limits_0^{\lambda_f}
 d\lambda'\frac{\langle
 0|\cos(\beta\phi)|k\rangle}{E_{k}(\lambda')-E_{0}(\lambda')},\nonumber
\eeq
where $|k\rangle$ is a short-hand notation  corresponding to the
soliton-antisoliton (breather) pair with momenta $k$ and $-k$ and
$E_{k}(\lambda)$ is the energy of such a pair. The expression for
the density of excited quasiparticles $n_{\rm ex}$ is therefore:
\beq\label{nex}
n_{ex}(\lambda_f)&\approx& \frac{1}{L}\sum_{k\neq 0} |\alpha_k(\lambda_f)|^2= \int\frac{d k}{2\pi}\left|\int_0^{\lambda_f} d\lambda'\langle
0|\partial_{\lambda'}|k\rangle\right|^{2}\nonumber\\
&=&4 \int\frac{d k}{2\pi}\left|\int_0^{\lambda_f} d\lambda'\frac{\langle
0|\cos(\beta\phi)|k\rangle}{E_{k}(\lambda')-E_{0}(\lambda')}\right|^{2}.
\eeq
Note that a factor of $2$  should appear here, due to the fact the
excited states correspond to pair of excitations, but it is exactly
canceled since we are counting each pair twice integrating over all
positive and negative momenta. Likewise in the same order of
approximation one can obtain the expressions for the heat density
$Q$:
\be
Q(\lambda_f)\approx 4 \int\frac{d k}{2\pi}
E_k(\lambda_f)\left|\int_0^{\lambda_f} d\lambda'\frac{\langle 0|\cos(\beta\phi)|k\rangle}{E_{k}(\lambda')-E_{0}(\lambda')}\right|^{2},
\label{qex}
\ee
and the entropy density:
\beq
\label{sd}
&&S_d\approx -{1\over 2}\int \frac{d k}{2\pi} |\alpha_k|^2\ln|\alpha_k|^2.
\eeq
There is an additional factor of $1/2$ in the expression
for the entropy coming from the fact that particles are excited only
in pairs and the contribution of each pair to the entropy (unlike
$n_{\rm ex}$) is not doubled.

As we argued earlier (see also Ref.~\cite{quench_short}) the
advantage of using intensive quantities like $n_{\rm ex}, S_d, Q$
over $P_{\rm ex}$ is that the regime of validity of the perturbation
theory for them is much bigger. Unlike for $P_{\rm ex}$ the quench
amplitude is not required to vanish as some power of the system
size. In Sec.~\ref{sec:exact} we will explicitly illustrate this
point for  two exactly solvable limits of the SG model: $\xi=1$ and
$\xi\ll 1$.

The adiabatic perturbation theory allows us to reduce the
computation of the dynamical response to static correlation
functions. The same is true if we are dealing with a slow processes
where the coupling changes gradually in time. In this case (see
Sec.~\ref{sec:gen}) there is an additional dynamical phase entering
the transition amplitude. For example for linear quenches
$\lambda(t)=\upsilon t$ instead of Eq.~(\ref{alpha}) we should use
(see also Refs.~\cite{nostro, kolkata, lincoln_book}):
\beq\label{alpha_slow}
 &&\alpha_k(\upsilon)=\int\limits_0^\infty
 d\lambda'\frac{\langle
 0|\cos(\beta\phi)|k\rangle}{E_{k}(\lambda')-E_{0}(\lambda')}\nonumber \\
&& \times \exp\left[{i\over\upsilon}\int_0^{\lambda'}
d\lambda'' \left(E_k(\lambda'')-E_{0}(\lambda'')\right)\right]
\eeq
in all expressions for $n_{\rm ex}$, $Q$ and $S_d$.

The matrix elements appearing in Eqs.~(\ref{alpha}) and
(\ref{alpha_slow}) are related to the form factors of the operator
$\cos(\beta\phi)$. In general the form factors represent the matrix
element of a particular operator between the vacuum and the
asymptotic states (eigenstates of the Bethe ansatz) created by the
Zamolodchikov-Faddeev operators corresponding to solitons,
antisolitons and breathers. Because the SG theory is
Lorentz-invariant it is convenient to use the rapidity variable
$\theta$, $-\infty<\theta<\infty$, which parametrizes the energy and
momentum of the soliton (breather):
\beq\label{1}
E=M_{s,B_{n}}\cosh\theta,\qquad  k=M_{s,B_{n}}\sinh\theta.
\eeq
The eigenstates of the SG model can then be labeled by
$|\theta_{n}\ldots\theta_{1}\rangle_{a_{n}\ldots a_{1}}$, where
$a_{i}=\{s,\bar{s},n=1\ldots[1/\xi]\}$ correspond to solitons,
antisolitons, and breathers. The correlation functions of an
arbitrary operator ${\cal O}(x,t)$ are written as an infinite series
expansion in terms of all the asymptotic states~\cite{essler_konik}:
\beq
&&\langle {\cal O}(x,t){\cal O}^{\dag}(0,0)\rangle
=\sum_{n=0}^{\infty}\sum_{\{a_{i}\} } \int\prod_{i=1}^{n}
\frac{d\theta_{i}}{(2\pi)^{n}n!}\nonumber\\
&&~~~\mathrm e^{i\sum_{j=1}^{n}k_{j}x-E_{j}t}\left|\langle0|{\cal O}(0,0)|\theta_{n}\ldots\theta_{1}\rangle_{a_{n}\ldots
a_{1}}\right|^{2}.
\eeq
The matrix elements entering Eqs.~(\ref{alpha}) and
(\ref{alpha_slow}) are then explicitly related to the form factors
of the operator $\cos(\beta\phi)$ in the two-particle asymptotic
states:
\be
\langle 0|\cos(\beta\phi)|k\rangle={1\over E(\theta)} {\langle
0|\cos(\beta\phi)|\theta,-\theta\rangle\rangle_{a_{2},a_{1}}}.
\label{matr_el}
\ee
Here the additional factor of $1/E(\theta)$ comes from the
relativistic normalization of the eigenstates parametrized by
$\theta$,
$\langle\theta|\theta'\rangle=\delta(\theta-\theta')/(2\pi)$. While in Eq.~(\ref{nex}) we
used a different normalization $\langle k|k'\rangle=\delta(k-k')/(2\pi)$. Taking into account the relations (\ref{1}) we indeed find that the normalization factor $1/|d_\theta k(\theta)|=1/E(\theta)$.

Let us discuss the two-soliton contribution to $n_{\rm ex}$ first, which
is expected to be dominant for $1\leq \xi<3$. The corresponding form
factors in the SG model are known in the literature~\cite{Smirnov,Lukyanov-MPL}:
\beq
&&F_{\exp(\pm
i\beta\phi)}(\theta=\theta_{2}-\theta_{1})\equiv\langle0|e^{\pm i\beta\phi}|\theta_{2},\theta_{1}\rangle_{s\bar{s}}
\nonumber\\
&&= {\cal G}_{\beta}G(\theta)\sum_{\sigma=\pm}
\frac{\cosh(\theta/2)\exp[\sigma\frac{\theta+i\pi}{2\xi}]}{\sinh(\frac{\theta+i\pi}{\xi})},
\label{ff}
\eeq
where we kept only the relevant factors (those which depend on
either $\theta$ or $\lambda_f$). We point out that the summation over
$\sigma$ comes from considering both the possibilities for a pair of
excitations (soliton-antisoliton and antisoliton-soliton). Also we
write both signs in the exponent of the operator $e^{\pm
i\beta\phi}$ since these form factors are invariant with respect to the
change of sign. Furthermore it is known that:
\beq
{\cal G}_{\beta}\sim M_{s}^{2\xi/(1+\xi)}.
\eeq
This result together with Eq.~(\ref{matr_el}) implies that $\langle
0|\cos(\beta\phi)|k\rangle\sim |M_s|^{\xi-1\over \xi+1}
P(\theta)\sim |\lambda_f|^{\xi-1\over 2} P(\theta)$, where $P(\theta)$
is some function which only depends on $\theta$. Next we use this scaling of the matrix element in  Eq.~(\ref{nex}) for finding $n_{\rm ex}$. If the integral over $k$ (or equivalently over $\theta$) converges at small $k\sim M_s$ then
we can perform the appropriate rescaling of the variables ($\lambda'
\to \lambda'/\lambda_f$ and $k \to k/M_s$ and send the upper limit of integration over $k$ to $\infty$. Then we immediately obtain the scaling $n_{\rm ex}\sim |M_s|\sim |\lambda_f|^\nu$, which
agrees with the general prediction (\ref{nex3}) in the case of a sudden quench.  A similar analysis of the matrix elements leads to the correct scaling for slow quenches characterized by an arbitrary exponent $r$ (see Ref.~\cite{nostro} for linear quenches).

To analyze the convergence of the integrals we need to find the
asymptotical behavior of the matrix element in Eq.~(\ref{nex}) at
large $\theta$. In order to do this, we look at the large $\theta$ behavior of
the $\theta$-dependent piece in Eq.~(\ref{ff}).
In particular we find:
\beq
G(2 \theta) \to \exp\left[\pm\frac{1}{2}\left(\frac{1}{\xi}+1\right)\theta\right],\qquad
\theta\rightarrow\pm\infty,
\eeq
which could be derived from the infinite-product-gamma-functions
representation of $G(\theta)$~\cite{Lukyanov-CMP} (or also, in a different way, see Eq.~(61) in the
book by F. Smirnov~\cite{Smirnov}). It is also straightforward to
deduce the asymptotics of the remaining part of the form factor to finally get:
\beq
F_{\cos(\beta\phi)}(2\theta)\sim \exp\left[\theta\left(\frac{3}{2}-\frac{1}{2\xi}\right)\right].
\eeq
To find the matrix element $\langle 0|\cos(\beta\phi)|k\rangle)$ we
need to divide this asymptotical form by the soliton energy (see
Eq.~(\ref{matr_el})) $E=M_s\cosh(\theta)\approx M_s
\exp[\theta]/2\approx k$ at large $\theta$ so that
\be
\langle 0|\cos(\beta\phi)|k\rangle\sim M_s^{\frac{\xi-1}{\xi+1}}
\exp\left[{\theta\over 2}{\xi-1\over\xi}\right]\sim M_s^{\frac{\xi-1}{\xi+1}} \left({E\over M_s}\right)^{{\xi-1}\over 2\xi}\!\!\!\!.
\label{mex_en}
\ee
This result correctly reproduces the scaling in the free fermionic
limit of the theory ($\xi=1$), which can be obtained by elementary
methods (see Sec.~\ref{sec:exact:fermi}). Such scaling implies a logarithmic singularity at $\xi=1$
for the heat density in a sudden quench $Q\sim
|\lambda_f|^2\log|\lambda_f|$ (see Eq.~(\ref{qex}), which is equivalent
to the log-divergence of the ground state energy. This divergence is
expected because for $\xi=1$ we have $\nu=(1+\xi)/2=1$, which
corresponds to the crossover point between the scaling $Q\sim
|\lambda_f|^{(d+z)\nu}$, expected for $\nu<2/(d+z)$, and the quadratic
scaling, expected for $\nu>2/(d+z)$. For $\xi>1$ the integral over
$k$ in Eq.~(\ref{qex}) diverges at large $k$, i.e. the main
contribution to the heat density comes from high energies (UV
limit), where the lowest form factors expansion is not justified and
we need to rely on the UV limit of the matrix elements discussed
below in Sec.~\ref{sec:cpt}. As for the other three quantities:
$n_{\rm ex}$, $P_{\rm ex}$, and $S_d$ the scaling of the matrix
element (\ref{mex_en}) together with Eqs.~(\ref{nex}) and (\ref{sd})
suggests that $n_{\rm ex}\sim |\lambda_f|^{d\nu}=|\lambda_f|^{1+\xi\over
2}$ holds for the whole soliton region $1\leq\xi<\infty$ and there
is no indication of crossover to the quadratic scaling at $\xi=3$
(corresponding to $\nu=2$). This result is however misleading because it
comes from the fact that the lowest order form factor expansion does
not correctly reproduce the ultra-violet asymptotics of the matrix
elements. As we mentioned above to get the right UV limit one needs to perform
an infinite re-summation of the form factors which implies that the
sum in Eq.~(\ref{p_ex}) should be taken over multiple soliton
states. This problem is well known in the equilibrium analysis of the UV
asymptotics of the static correlation functions~\cite{KL}. In the
next section we will show how the power two emerges from analyzing
Eq.~(\ref{p_ex}) using the conformal perturbation theory.

Likewise one can analyze the scaling of the matrix elements in the
breather region ($\xi<1$). Since we have $\nu=(1+\xi)/2<1$ we expect
that the scaling of all the quantities we are interested in $Q,
P_{\rm ex}, n_{\rm ex}, S_d$ is dominated by low energies where the
lowest form factor expansion is justified. So for $\xi<1$ we need to
repeat the analysis above with the only difference that the states
$|k\rangle$ in Eq.~(\ref{alpha})  denote pairs of breather states
with opposite momenta. The form factors for breathers have a
different scaling form than those for solitons. With the operator
$\cos(\beta\phi)$ it is possible to have non-zero form factors of a
single breather $B_{n}$ of the form
$\langle0|\cos(\beta\phi)|B_{n}\rangle$ if $n$ is even. Those form
factors do not depend on the rapidity at all
(see~\cite{toliacoupled} for more details). The first
$\theta$-dependent contribution into Eq.~(\ref{p_ex}) comes from
exciting two $B_1$ breathers. These breathers are the only addition
to the solitons in the spectrum for $1/2\leq \xi< 1$ and they are
also the only surviving excitations in the noninteracting bosonic
limit of the SG model $\xi\to 0$. Therefore here we will concentrate
only on the scaling analysis of $n_{\rm ex}$ and other observables
coming from exciting pairs of $B_1$ breathers with opposite
momenta~\endnote{One can show the additional contribution coming
from the zero $B_2$ breathers in the regime $0<\xi<1/2$  (see
Ref.~\cite{gritsev:quench}) does not change the universal scaling of
the quantities of interest.}. The matrix element describing this
process is found through the corresponding form factors which were
computed in Ref.~\cite{toliacoupled} using bootstrap, the procedure
of fusion of several breathers~\cite{Zamolodchikovs79}:
\beq
\langle 0|\cos(\beta\phi)|k\rangle&=&{1\over E(\theta)}\langle 0|\cos\beta\phi|B_{1}(\theta)B_{1}(-\theta)\rangle\nonumber\\
&=&{1\over 2M_s\cosh\theta}\mathcal G_\beta \gamma^2 {\sin(\pi\xi)\over \pi \xi}R(2\theta),
\label{matr_el1}
\eeq
where
\be
\gamma=2\cos\left[{\pi\xi\over 2}\right]\sqrt{2\sin\left[{\pi\xi\over 2}\right]}\exp\left[-\int_0^{\pi\xi} {dt\over 2\pi}{t\over \sin t}\right]
\ee
and
\begin{widetext}
\be
R(\theta)=\exp\left[4\int_0^\infty {dt\over t}{\sinh
(t)\sinh(t\xi)\sinh (t(1+\xi))\over \sinh^2(2t)}\right]
\exp\left[8\int_0^\infty {dt\over t}{\sinh (t)\sinh(t\xi)\sinh
(t(1+\xi))\over \sinh^2(2t)}\sinh^2 t\left(1-i{\theta\over
\pi}\right)\right].
\ee
\end{widetext}
Since the dependence of the matrix element (\ref{matr_el1}) on $M_s$
(and hence on $\lambda_f$) is again found via the function ${\cal
G}_{\beta}$, we conclude that the scaling of $n_{\rm ex}$ (as well
as $Q, S_d$ and $P_{\rm ex}$) on $\lambda_f$ is the same as in the
repulsive regime, e.g. $n_{\rm ex}\sim |\lambda_f|^{(1+\xi)/2}$. We only
need to verify that the integrals over momenta $k$ converge at large
$k$. In order to do this we need to know the  asymptotics of the
function $R(2\theta)$ at large $\theta$.  The  scaling at large
rapidity for the function $R(\theta)$ (defined in
\cite{Lukyanov-MPL}) is $R(\theta)\sim F_{B_{1}}^{2}$, which does
not depend on $\theta$. This saturation, which is the consequence of bootstrap, is actually true for higher
order form factors as well. For example $\langle
0|\cos\beta\phi|B_{2}(\theta)B_{2}(-\theta)\rangle \sim
(F_{B_{2}})^{2}$. On the other hand, because of the bootstrap
relation, this should be proportional to $\langle
0|\cos\beta\phi|B_{4}\rangle$ which is a finite number. The
one-breather form factors of the type $\langle
0|e^{i\beta\phi}|B_{n}(0)\rangle\equiv F_{B_{n}}$ can be computed
from the residue of the soliton-antisoliton form factors at points
$\theta_{n}=i\pi(1-n\xi))$, or, equivalently from the bootstrap (the
definition of $F_{B_{n}}$ can be found  in \cite{toliacoupled}). For
example, the breather $B_{2}$ is a bound state of two breathers
$B_{1}$. Using this bootstrap procedure we can find
\be\label{1B}
F_{B_{n}}^{\exp(i\beta\phi)}\!\!=\!\frac{{\cal
G}_{\beta}\sqrt{2}\cot(\frac{\pi\xi}{2})\sin(\pi
n\xi)\exp[I(-\theta)]e^{\frac{i\pi
n}{2}}}{\sqrt{\cot(\frac{\pi\xi
n}{2})\prod_{s=1}^{n-1}\cot^{2}(\frac{\pi \xi s}{2})}},
\ee
where
\beq\label{I}
I(\theta)=\int_{0}^{\infty}\frac{dt}{t}\frac{\sinh^{2}
\left[t\left(1-\frac{i\theta}{\pi}\right)\right]\sinh[t(\xi-1)]}{\sinh[2t]\cosh[t]\sinh[t\xi]}.
\eeq

\begin{figure}[h]
\includegraphics[width=6cm,angle=0]{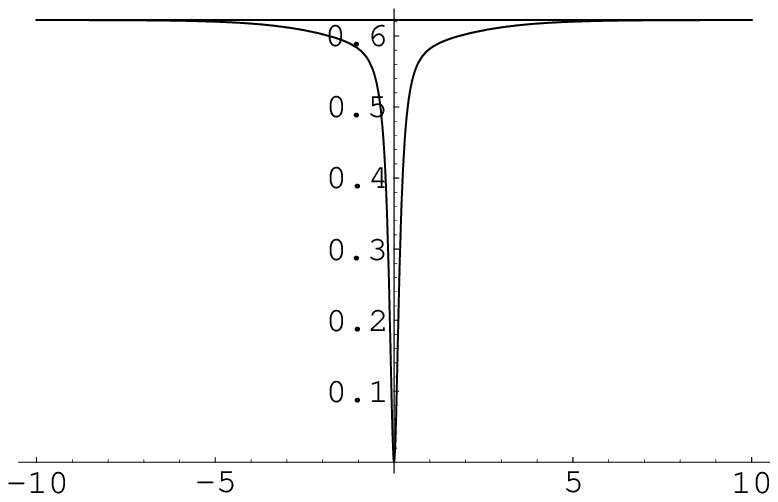}
\includegraphics[width=6cm,angle=0]{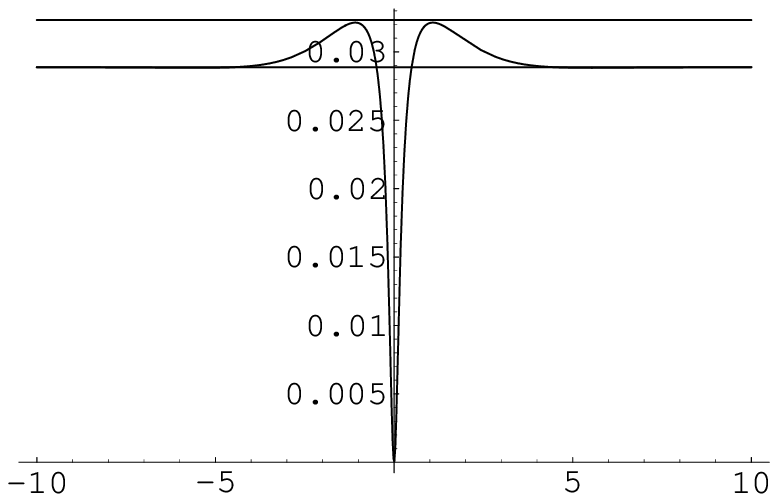}
\caption{$B_{1}B_{1}$ (top) and $B_2B_2$ (bottom) breather form factor for $\xi=1/41$ as a
function of the rapidity $\theta$. The saturation lines are given by
$(F_{B_{1}})^{2}$ and $(F_{B_{2}})^{2}$ respectively.}
\label{b1-b1}
\end{figure}

The saturation of the form factors implies that the matrix elements
$\langle 0|\cos(\beta\phi)|k\rangle$ behave as $1/E(k)$ at large $k$
ensuring the convergence of integrals in Eqs.~(\ref{nex}),
(\ref{qex}), and (\ref{sd}) and thus the validity of the universal
scalings. In the next section using the conformal perturbation
theory we will also show that there are no UV issues with these
scaling relations, thus the lowest form factor expansion is indeed
justified. We note that the saturation of the breather form factors
can be also understood from the fact that $\cos(\beta\phi)$
corresponds to the trace of the stress-energy tensor, whereas
breathers correspond to the field $\phi$ itself. This saturation is
thus expected for all breather contributions corresponding to
exciting more than a single pair of breathers. In Fig.~\ref{b1-b1}
we show the dependence of the $B_1B_1$ and $B_2B_2$ breather form
factors as a function of the rapidity. Both plots indicate clear
saturation with $\theta$ at the value corresponding to the square of
the single breather form factor.

\section{Analysis of the high energy excitations based on the fidelity susceptibility and the conformal perturbation theory}
\label{sec:cpt}

The other approach for finding the scaling of $P_{\rm ex}$ and other
quantities with the quench amplitude $\lambda_f$ is based on the analysis of the
fidelity susceptibility $\chi_f$, which is in turn related to the
correlation functions. From Eq. (\ref{pex_ups}) we know that for
sudden quenches of  small amplitude we can write:
\be
P_{\rm ex}(\lambda_f)\approx L|\lambda_f|^2\chi_f(\lambda_f),
\label{pex1}
\ee
with $\chi_f$  defined in Eq.~(\ref{chif}). A more accurate analysis based on the Cauchy-Schwartz inequality
(see Sec.~\ref{sec:geom_tens}) shows that $P_{\rm ex}$ is bounded by the integral of $\chi_f$ as
$P_{\rm ex}(\lambda_f)\leq L\lambda_f \int_0^{\lambda_f} d\lambda \chi_f(\lambda)$, still the scaling of $\lambda_f\int_0^{\lambda_f} \chi_f(\lambda)d\lambda$ and $\lambda_f^2
\chi_f(\lambda_f)$ is the same for any power-law dependence of $\chi_f(\lambda_f)$, which we typically find near the critical point. Moreover in this section we will be primarily interested in the UV
limit, corresponding to exciting high energy states, where
Eq.~(\ref{pex1}) is justified by the conformal perturbation theory.
Likewise one can obtain the expression for heat density:
\be
Q \approx |\lambda_f|^2\chi_E(\lambda_f),
\label{Q}
\ee
with :
\be
\label{chie}
\chi_E(\lambda)={1\over L^d}\sum_{n\neq 0} { |\langle 0|V|n\rangle|^2\over E_n(\lambda)-E_0(\lambda)}.
\ee
The susceptibility $\chi_E$ is equal to the generalized adiabatic susceptibility of order one: $\chi_E=\chi_1$ according to Eq.~(\ref{chim}).

For sudden quenches the wave function does not change during the
quench. Hence the total energy after the quench is given by the
expectation value of the new Hamiltonian taken with the initial
wavefunction:
\be
E_+=\langle \Psi_0| H_0+\lambda_f V|\Psi_0\rangle,
\label{en}
\ee
where $H_0$ is the free conformal Hamiltonian corresponding to the
critical point and $V=-2\int dx\cos(\beta\phi)$. In the conformal
limit the ground state is invariant under the uniform phase rotation
$\phi(x)\to\phi(x)+\phi_0$, where $\phi_0$ is an arbitrary phase.
Using this symmetry the second term in Eq.~(\ref{en}) drops out and
thus the total energy does not change during the quench. The heat is
defined as the difference between the total energy and the adiabatic
(ground state) energy of the instantaneous Hamiltonian ($E_{\rm
gs}(\lambda_f)$). Therefore we see that in our case the heat is equal
by magnitude and opposite by sign to the change of the ground state
energy: $Q=-(E_{\rm gs}(\lambda_f)-E_{\rm gs}(0))$. One can recognize
that Eq.~(\ref{Q}) indeed gives (up to the sign) the second order
perturbative correction to the ground state energy. As in the case
of $P_{\rm ex}$, the perturbative expression is justified if we are
interested in understanding the UV limit, where the energy
denominator (see Eq.~(\ref{chie})) is large.

As we argued in Sec.~\ref{sec:gen} the susceptibilities $\chi_E$ and
$\chi_f$ can be expressed through the integral of the connected part
of the imaginary time correlation function $G(x,\tau)$:
\beq
G(x,\tau)&=&\langle 0|\cos(\beta\phi(x,\tau))\cos(\beta\phi(0,0))|0\rangle\nonumber\\
&-&\langle 0|\cos(\beta\phi(0,0))|0\rangle^2.
\eeq
Namely, for a translationally invariant system we have:
\beq
\chi_E&=&\int_0^\infty d\tau\int_{0}^{L} dx\, G(x,\tau),\label{chi_e}\\
\chi_f&=&\int_0^\infty d\tau\int_{0}^{L} dx\, \tau
G(x,\tau).\label{chi_f}
\eeq
Let us point out that $G(x,\tau)$ is a monotonically decreasing
non-negative function of $r=\sqrt{x^2+\tau^2}$ vanishing in the
limit $r\to\infty$. The non-negativity of $\int dx \, G(x,\tau)$ follows
from the Lehmann's representation:
\be
\int_{0}^{L} dx \, G(x,\tau)={1\over L}\sum_{n\neq 0} |\langle 0|
V|n\rangle|^2\exp[-\tau(E_n-E_0)].
\label{Lehmann}
\ee
The relativistic invariance of the SG model also implies the non-negativity of
$G(x,\tau)$ itself.

The scaling of $P_{\rm ex}$ and $Q$ with $\lambda_f$, at small
$\lambda_f$, is thus related to the small $\lambda_f$ behavior of the
integrals of $G(x,\tau)$. The long distance behavior of the
correlations functions at scales $r\gg M_s^{-1}$ is given by the form factor
expansion and can be analyzed by the methods of the previous section,
leading to essentially the same predictions. In particular, because at
any finite $\lambda_f$ the spectrum of the SG model is gapped,
$G(x,\tau)$ exponentially decays on a correlation length scales
$r\approx \zeta\sim 1/M_s~\sim 1/|\lambda_f|^{(1+\xi)/2}$. The scaling
dimension of the operator $\cos(\beta\phi(x,\tau))$, entering the
correlation function, is $\beta^2/4\pi=2\xi/(1+\xi)$,  implying that
the scaling dimension of $G$ is twice that. Thus we can
estimate the contribution to the scaling of $\chi_f$ coming from
$\tau,x~\sim \zeta$ as $\chi_f\sim \zeta^3/\zeta^{4\xi/(1+\xi)}\sim
|\lambda_f|^{(1+\xi)/2-2}\sim |\lambda_f|^{\nu-2}$. This is precisely the scaling we obtained
from the form factors expansion.

The primary goal of this section is to analyze the UV (short
distance) limit where the form factor expansion in the lowest order does not
reproduce the correct scaling behavior, as we pointed out in Sec.~\ref{sec:FF}. For the contribution to the
susceptibilities and  thus to $P_{\rm ex}$ and $Q$ (as well as to
$n_{\rm ex}$ and $S_d$) coming from $r\ll \zeta\sim M_s^{-1}$ one
can use the cosine term as a small perturbation over the free (or
conformal) limit. We note that by the uncertainty principle the short
distances correspond to  exciting states with
high energies (and momenta) $E_n\gg M_s$ ($k\gg 1/\zeta$). In turn
the short distance behavior of the correlation functions is governed by
power-law asymptotics with the exponents given by the scaling
(conformal) dimensions of the operator $\cos(\beta\phi)$, (which is
$\beta^{2}/4\pi=2\xi/(\xi+1)$ as previously said). Thus in Euclidean notations:
\be
G(x,\tau)\sim
\frac{1}{(x^{2}+\tau^{2})^{\beta^{2}/4\pi}}.
\label{gxtau}
\ee
While this result is natural, it is somewhat nontrivial to obtain
from the form factor perturbation theory since it requires an
infinite re-summation of the form factors as it is nicely
demonstrated in Refs.~\cite{Mussardo-book, BK}. So we can conclude
that in order to correctly reproduce the high energy contribution to
$P_{\rm ex}$ and $Q$ it is not sufficient to restrict the excited
states in Eqs.~(\ref{chif}), (\ref{p_ex}) and (\ref{chie}) to simply
pairs of solitons and antisolitons with opposite momenta. It appears
that only in the free fermionic point $\xi=1$ the lowest
soliton-antisoliton contribution to $G(x,\tau)$ correctly predicts
the exponent in Eq.~(\ref{gxtau}). We believe that in general in our
out-of-equilibrium setup it is possible to use a combination of some
RG-type computations with the form factors based approach similar to
the one in Ref.~\cite{CET}.

From Eqs.~(\ref{chi_e}) and ~(\ref{chi_f}), we see that at the
critical point $\chi_E$ has IR divergence at $\beta^2\leq 4\pi$
corresponding to $\xi<1$ and $\chi_f$ has such divergence at
$\beta^2\leq 6\pi$ corresponding to $\xi<3$. By now we understand
that this divergence simply implies that the scaling of
$Q(\lambda_f)$ and $P_{\rm ex}(\lambda_f)$ is non-analytic in
$\lambda_f$ or the system size in these domains. This divergence
also indicates that the main contribution to the susceptibilities,
and thus to the heat and the probability of exciting the system,
comes from small energies $E\lesssim M_s$, where one can reliably
use the lowest form factors expansion. Conversely for $\xi>1$ the
susceptibility $\chi_E$ has a UV divergence, which is cutoff by the
non-universal short distance regularization of the SG model.
Similarly for $\xi>3$ the fidelity susceptibility $\chi_f$ is
dominated by short distances or high energies. This implies that in
these regimes of UV divergence, $\chi_E(\lambda)$ and
$\chi_f(\lambda)$ approach non-universal constant values as $\lambda
\to 0$ (note that these values are strictly positive because
$G(x,\tau)$ is strictly positive). Therefore the leading
asymptotical behavior for the heat becomes quadratic for $\xi>1$:
$Q\sim |\lambda_f|^2\chi_E(0)$. Likewise the scaling  for $P_{\rm
ex}$ becomes extensive for $\xi>3$: $P_{\rm ex}\sim
L\lambda_f^2\chi_f(0)$. This exactly agrees with the crossover to
the quadratic scaling anticipated in the Sec. \ref{sec:gen}, that occurs for
$(d+z)\nu>2$ for $Q$ and for $d\nu>2$ for $P_{\rm ex}$ . This
crossover is directly analogous to the crossover between
non-analytic and analytic regimes of scaling for the linear quenches
in higher dimensions~\cite{tolianature}. Similar crossover between
perturbative and non-perturbative scalings at $\xi=3$ occurs for the
density of excitations and the entropy.

While the short distance behavior of the correlation function given
by the conformal limit (\ref{gxtau}) is intuitively anticipated,
finding corrections to this scaling due to the finite value of the
coupling $\lambda$ requires using conformal perturbation theory ~\cite{zamo_}. In
the latter the starting point is the conformal field theory (free
boson compactified on a circle of radius $R$ in our case). The
Hilbert space then is the Verma module - representation space of the
Virasoro algebra. The different basis states in this approach are
labeled by  two integers $(n,m)$~\footnote{In the Coulomb gas picture
these integers correspond to magnetic and electric charges.}
corresponding to the eigenvalues $p_{\pm}=n/R\pm m/2R$ of the
momenta of the left/right movers
$\varphi(z),\bar{\varphi}(\bar{z})$, with $\Phi(x,t)=\varphi(z)+
\bar{\varphi}(\bar{z})$. The perturbed theory is constructed by
adding vertex operators
$V_{(n,m)}(z,\bar{z})=\exp[i(p_{+}\varphi(z)+p_{-}\bar{\varphi}(\bar{z}))]$
to the free theory. The scaling dimensions of the chiral parts of
these vertex operators are $\Delta_{\pm}=p_{\pm}^{2}/2$ so that the
scaling dimension $h$ of $V_{(n,m)}$ is $h=\Delta_{+}+\Delta_{-}$.
The sine-Gordon theory (\ref{sgham}) is considered then as a
perturbation of this free theory by the vertex operators with
$n,m=\{\pm 1,0\}$.
\beq
&&H_{SG}=H_{CFT}+V,\\
&&V=-2\lambda_f\int_{0}^{L}(V_{(1,0)}(z,\bar{z})+V_{(-1,0)}(z,\bar{z}))dx\nonumber\\
& &\Phi(x,t)=\sqrt{4\pi}\phi(x,t),\quad R=\sqrt{4\pi}/\beta^{2}.
\eeq
The scaling dimension of $V$ is then $\beta^{2}/4\pi$. Since the
Hilbert space is built by the action of the vertex operators on the
vacuum state, $|n,m\rangle=V_{(n,m)}(0,0)|vac\rangle$,
(higher-energy states are built by the application of powers of the
chiral components of the Kac-Moody currents $\partial\varphi$ to
these states) the conformal perturbation theory is simply a
perturbation theory in this space, and the effect of the
perturbation is an expansion in powers of the $(M_{s}L)^{2-h}$,
where $h=\beta^{2}/4\pi$ is the scaling dimension of the
perturbation (so this expansion is in integer powers of $\lambda_f$).
In Ref.~\cite{KL} such expansion to the third power of $\lambda_f$ was
applied to the correlation function. In particular, it was found
that
\be
\langle 0|\mathrm e^{i\beta\phi(x,\tau)}\mathrm
e^{-i\beta\phi(0,0)}|0\rangle= {1\over
r^{\beta^2/(2\pi)}}+C\lambda_f^2 {1\over r^{\beta^2/\pi-4}}+\dots,
\label{corr_func}
\ee
where $r=\sqrt{x^2+\tau^2}$ and $C$ is a constant. In the repulsive
regime $4\pi<\beta^2<8\pi$ where this expansion was derived the
second term clearly decays slower at large $r$ than the first
leading term indicating that the Taylor expansion in $\lambda_f$ can
break down. This is of course anticipated because $\cos(\beta\phi)$
is the relevant perturbation. Conversely at small $r$ the first term
corresponding to the conformal limit is dominant.

If we use this expansion to compute $\chi_E$ according to
Eq.~(\ref{chi_e}) then as we discussed above for $\beta^2<4\pi$ (or
$\xi<1$) the first term gives IR divergence which should be fixed by
using the proper long distance asymptotics of the correlation
function. On the other hand for $4\pi<\beta^2<6\pi$ ($1<\xi<3$) the
first term gives a finite constant contribution to $\chi_E$ while
the second term proportional to $\lambda_f^2$ gives a divergence which
again has to be fixed by the proper long distance asymptotics. For
$3<\xi<5$ the second term gives a convergent contribution indicating
that $\chi_E$ has analytic expansion in $\lambda_f$ up to $\lambda_f^2$
but the following term  $\propto \lambda_f^4$ is expected to give a non-analytic
contribution and so on. Thus from the conformal perturbation we can
anticipate that the asymptotic expansion for the heat in powers of
$\lambda$ has the following structure
\be
Q=\sum_{0<n<[(1+\xi)/2]} \tilde C_n \lambda_f^{2n} + \tilde B |\lambda_f|^{1+\xi}+o(|\lambda_f|^{1+\xi}),
\label{qexp}
\ee
where $\tilde C_n$ are some non-universal constants which depend on
the short distance cutoff and $\tilde B$ is a universal constant
determined by the low energy (long distance) contribution to
$\chi_E$ (and thus to the heat). At the special points $1+\xi=2n$ the
constant $\tilde B$ is expected to have logarithmic divergence.

Similar analysis can be done for $\chi_f$ using
Eq.~(\ref{chi_f}) and the expansion (\ref{corr_func}). Now we
observe that the first conformal contribution diverges at
$\beta^2<6\pi$ ($\xi<3$). As before this divergence is cutoff by
using the proper long distance asymptotics of $G(x,\tau)$ and
results in the non-analytic scaling of $\chi_f(\lambda)$. For
$6\pi<\beta^2<7\pi$ ($3<\xi<7$) the first term in
Eq.~(\ref{corr_func}) gives a constant contribution to $\chi_f$,
while the second term diverges and thus results in a  non-analytic
contribution. For $7<\xi<11$ we see that both the first and the
second terms in (\ref{corr_func}) give convergent contribution to
$\chi_f$, but we can anticipate that the third term gives a
non-analytic correction and so on. We can therefore anticipate
the scaling for the probability of exciting the system:
\be
P_{\rm ex}=L\sum_{n=1}^{(1+\xi)/4} \breve{C}_n \lambda_f^{2n} + \breve{B} \lambda_f^2 L^{4/(1+\xi)} +o(|\lambda_f|^2 L^{4/(1+\xi)}),
\ee
where as before $\breve{C}_n$ are non-universal constants and
$\breve{B}$ is universal and also expected to have log-divergence at
the special points $1+\xi=4n$. While we can expect a similar structure
of the expansion for $n_{\rm ex}$ and $S_d$, it is harder to analyze
these quantities in the UV regime because they do not have an explicit
representation through the correlation functions. We simply note that the
crossover to the leading quadratic scaling for these quantities when
$\xi>3$ (with probable log corrections for the entropy) is expected
on general grounds as a manifestation of the validity of usual
perturbation theory applicable to the high energy excitations, where the
probability of the transition to be excited scales as
$|\alpha_n|^2\sim\lambda_f^2$. Since the low energy excitations coming
from creating soliton-antisoliton pairs give $n_{\rm ex}, S_d\sim
|\lambda_f|^{(1+\xi)/2}$ we expect that the high energy excitations
will dominate precisely when $(1+\xi)/2>2$.

One can actually analyze the energy of an arbitrary state
$|\Psi\rangle$, obtained by acting with the operator $\Psi$ on the vacuum
$|\Psi\rangle= \Psi |0\rangle $, as a function of $\lambda_f$ using
the nonlinear integration equation approach~\cite{DDV, FRT}. The
corresponding expansion in powers of $\lambda_f$ reads:
\be
\frac{E_{\Psi}}{M_{s}}=-\frac{\pi c}{6l}+\frac{2\pi
h_{\Psi}}{l}+Bl+\sum_{n=1}^{\infty}C_{n}(\Psi)l^{n(2-h_{\Psi})}
\label{epsi}
\ee
where $l=M_{s}L$, $B=-\tan(\pi\xi/2)/4$ is the bulk energy in the
thermodynamic limit, $c$ is the central charge (equal to 1 in our
case), $C_{n}(\Psi)$ are non-universal constants. The bulk energy
becomes infinite at the point where $\xi$ is an odd integer. At this
points there is a value of $n$ for which $n(2-h)=1$, i.e.
$\beta^{2}=8\pi(1-1/(2n))$ and $C_{n}\rightarrow\infty$. The
infinite contributions from the bulk part and from the "C-part"
cancel giving the logarithmic singularity proportional to
$l\log(l)$. These expectations are consistent with the findings of
Refs.~\cite{DDV}, \cite{FRT}. In particular, it was found that at the
special points $\beta^{2}=8\pi(1-1/2n)$, corresponding to
$\xi=2n-1$, the expression for the ground state energy gives
logarithmic singularities. Exactly at these points the
short distance expansion method of Konik and LeClair \cite{KL} gives
the divergency of the $n$-th order of the perturbative expansion.
Noting that for sudden quenches the heat is related to the ground
state energy and thus to $\chi_E$ we see that the expansion
(\ref{epsi}) is consistent with the asymptotical form (\ref{qexp})
suggested above.

Note that in the attractive regime $\beta^2<4\pi,\;\xi<1$ the UV
asymptotics of the correlation functions always give long distance
divergent contribution to both $\chi_E$ and $\chi_f$ indicating that
the corresponding susceptibilities are determined by the IR region
where the lowest form factors expansion is accurate. This confirms
that the general scalings obtained in Ref.~\cite{quench_short} (see
also Sec.~\ref{sec:gen}) and verified in the previous section are
justified in the whole breather region.

Let us point out that a very similar analysis can be extended to slow
quenches with arbitrary exponent $r$ (see Eq.~(\ref{lambda_t})). The
validity of the perturbative quadratic and extensive scaling of $P_{\rm
ex}$ is determined by the convergence of the adiabatic
susceptibility of order $2r+2$ (see Eqs.~(\ref{pex_ups}) and
(\ref{chim})). In particular, for linear quenches, the relevant
susceptibility $\chi_4(\lambda)$ is always IR divergent in the
regime of our interest, $0<\beta^{2}<8\pi$, implying that the
scaling of $P_{\rm ex}$ in the case of linear adiabatic quenches in
one dimension~\cite{nostro} remains  non-analytic in the whole
gapped regime of the SG-model and the form factors analysis in this
range is always justified. Thus in order to analyze the crossover
between analytic and nonanalytic regimes of scaling we need to
consider the generalizations of the SG model to higher dimensions.
This can be easily done in the two limits $\xi=1$ and $\xi\ll 1$
corresponding to free massive fermionic and bosonic theories
respectively. We will discuss these generalizations and the
crossovers in the next section.

\section{Exactly solvable cases: free massive bosons and fermions}
\label{sec:exact}

In this section we will focus on two limits of the SG model where
the dynamics can be analyzed exactly without the need to rely on the
adiabatic perturbation theory or other approximations. These limits
have also the advantage to have straightforward multi-dimensional
generalization. First we consider the limit of $\beta\ll 1$. In this
case the cosine term in the Hamiltonian~(\ref{sgham}) can be
expanded to the quadratic order in the phase $\phi$ and the problem
maps into the free scalar massive bosonic field theory. The quench
process corresponds to a sudden increase of the mass starting from
zero. The quench dynamics in this regime was recently considered  in
Ref.~\cite{cazalilla_quench} from a different point of view, analyzing
 the correlation functions and the asymptotic steady state. The other solvable limit corresponds to
$\beta=2\sqrt{\pi}$ ($\xi=1$). For this value of $\beta$ the problem
deals with hard-core bosons or non-interacting fermions. In the
language of fermions the quench process corresponds to a  sudden
turn on of a commensurate periodic potential, which opens a gap at
the fermi momentum. In the following two sections we will discuss
these limits in detail, analyze the scalings of the density of
quasiparticles, heat, and entropy and show how those results agree
with the general scaling predictions of Sec. \ref{sec:gen}. For
these solvable cases  in Sec. \ref{sec:temp} we will be able also to
extend the  exact results to  quenches at finite temperatures.

\subsection{Elementary derivation of the scaling relations using the adiabatic perturbation theory}

Before proceeding with the analysis of specific bosonic and
fermionic models we show how the scaling relations and the crossover
to the quadratic scaling can be understood if we are dealing with a
free theory. This discussion goes along the lines with that of
Ref.~\cite{toliapoint} for slow linear quenches. Using the general
discussion of Sec. \ref{sec:gen} we find that within the adiabatic
perturbation theory the amplitude for creating an excited state in a
sudden quench is:
\be
\alpha_n(\lambda_f)= \int_0^{\lambda_f} d\lambda\; {\langle n|V|0\rangle\over E_n(\lambda)-E_0(\lambda)}.
\ee
If we are dealing with a free spatially uniform Hamiltonian then the
operator $V$ must be quadratic in the creation and annihilation
operators and conserve the total momentum. Thus acting on the vacuum
state $V$ can either leave it intact or create a pair of
quasiparticles with opposite momenta. Hence all the intermediate
states can be characterized by the momentum label $q$. Thus the
total probability to excite the system is:
\be
P_{\rm ex}=1-\prod_{q >0} (1-p_{\rm ex}(q)),
\ee
where  the product is taken over different pairs of two-particle
states with opposite momenta and $p_{\rm
ex}(q)=|\alpha_q(\lambda_f)|^2$ is the probability to create the
pair. In the limit when $\sum_{q>0} p_{\rm ex}(q)\ll 1$ the product
can be approximated by the sum and we have:
\be
P_{\rm ex}\approx \sum_{q>0} p_{\rm ex}(q).
\ee
Note that since each term in the sum corresponds to a pair of
quasiparticles the number of excited quasiparticles is given by the
same expression above multiplied by a factor of two, which can be
absorbed into the sum by summing over all positive and negative
momentum states. Thus
\be
n_{\rm ex}\approx {1\over L^d}\sum_q \left|\int_0^{\lambda_f} d\lambda\; {\langle q|V|0\rangle\over E_q(\lambda)-E_0(\lambda)}\right|^2,
\label{nex_pert}
\ee
where $|q\rangle$ is a short-hand notation for the state with two
excited quasiparticles with momenta $q$ and $-q$ and  $E_q(\lambda)$
is the corresponding excitation energy of the pair. Let us analyze
the expression above using the scaling arguments (see also
Refs.~\cite{toliapoint, kolkata, lincoln_book}). Near the critical
point we expect that
$E_q(\lambda)-E_0(\lambda)=|\lambda|^{z\nu}F(q/\lambda^\nu)$, with
$F(x)$ a scaling function. In the limit $x\gg 1$ we must have
$F(x)\sim x^z$ to ensure the scale invariance at the critical point.
Similarly we  write the scaling of the matrix element:
\be
{\langle q|V|0\rangle\over E_q(\lambda)-E_0(\lambda)}={1\over |\lambda|}G (q/|\lambda|^\nu),
\ee
where $G(x)$ is another scaling function. The same requirement of
scale invariance at the critical point gives that $G(x)\sim
1/x^{1/\nu}$ at $x\gg 1$. Some examples where the scalings and
asymptotics for the matrix element and the energy were explicitly
confirmed can be found in Refs.~\cite{kolkata, tolianature}. Using
these scaling results it is straightforward to find the scaling for
the density of excited quasiparticles. In particular, focusing on
the limit $|\lambda_f|\gg L^{-1/\nu}$, where the summation over
momenta can be replaced by the integration, and changing variables
$q=|\lambda_f|^\nu \eta$ and $\lambda=\lambda_f\xi$ in
Eq.~(\ref{nex_pert}) we find that
\be
n_{\rm ex}\approx |\lambda_f|^{d\nu}\int {d^d\eta\over (2\pi)^d} \left|\int_0^{1} {d\xi \over  \xi}\; G(\eta/|\xi|^\nu)\right|^2.
\label{nex_pert1}
\ee
A similar expression is valid for the heat density:
\be
Q\approx |\lambda_f|^{(d+z)\nu}\int {d^d\eta\over (2\pi)^d} F(\eta)\left|\int_0^{1} {d\xi \over  \xi}\; G(\eta/|\xi|^\nu)\right|^2.
\label{qpert}
\ee
We thus see that we are reproducing the scaling prediction of Sec.
\ref{sec:gen}, Eqs.  (\ref{nex3}) and (\ref{Qex3}), for sudden
quenches ($r=0$) as long as the limits of the integration over
$\eta$ can be extended to infinity. This is only justified if the
integral over $\eta$ converges at large $\eta$. Using the
asymptotical expression for the scaling functions $G$ and $F$ we see
that this is the case for $d\nu<2$ if we are analyzing $n_{\rm ex}$
and for $(d+z)\nu<2$ if we are analyzing the heat. Alternatively the
integrals over the momenta are dominated by the high energy cutoff
and the scalings of the corresponding quantities become quadratic
with $\lambda_f$, in accord with the predictions of Sec.
\ref{sec:gen}. A similar analysis shows that the adiabatic
perturbation theory reproduces the correct scaling also for the
entropy density $S_d\sim |\lambda_f|^{d\nu}$ when $d\nu<2$. The same
approach confirms the correct scaling of all these quantities for
small quench amplitudes $|\lambda_f|\ll 1/L^{1/\nu}$, e.g. $n_{\rm
ex}, P_{\rm ex}\sim |\lambda_f|^2 L^{2/\nu-d}$. The easiest way to
see this is from matching the quadratic scaling and the universal
scaling when the correlations length $\xi\sim 1/|\lambda_f|^\nu$
becomes of the order of the system size. In the next two sections we
will explicitly show how these crossovers emerge from the two free
limits of the SG model.

\subsection{Free massive bosons}

Consider the SG Hamiltonian (\ref{sgham}) when $\beta \to 0$. In
this limit we can Taylor expand the cosine term up to the quadratic
order in the field $\phi$ and obtain a quadratic Hamiltonian, that
in the Fourier space has the form:
\be\label{SGqua}
H = {1\over 2}\sum_q  | \Pi_q|^2 +\kappa_q(\lambda)|\phi_q|^2,
\ee
with $\kappa_q(\lambda)=q^2+2\lambda(t)\beta^2$ and
$\Pi_q^\dagger,\phi_q^\dagger=\Pi_{-q},\phi_{-q}$. We point out that
here as in Eq.~(\ref{sgham}) we have set $\hbar v_s=1$, with $v_s$
being the sound velocity. For each momentum $q$ the fields
$\Pi_q^\dagger$ and $\phi_q$ are conjugates,
$[\Pi_q^\dagger,\phi_{q'}]=i\delta_{q,q'}$. Therefore we are dealing
with a sum of independent harmonic oscillators with a time-dependent
compressibility $\kappa$ (or the inverse mass). The wave function
for this time-dependent Hamiltonian can be found exactly, for
instance in Refs.~\cite{tolianature, kolkata} this was done for the
case of an adiabatic-linear quench ($\lambda(t)=\upsilon t$). Here
we focus on the case of a sudden quench and show that the exact
results agree with the general scaling predictions  that have been
presented in the Sec.\ref{sec:gen}. Furthermore the Hamiltonian
(\ref{SGqua}) can be easily extended to higher dimensions and thus
we are able to see how the crossover to the quadratic scalings and the usual
perturbation theory emerges as the corresponding exponents $d\nu$ or
$(d+z)\nu$ exceed two. We note that in this limit of the SG model
$\nu=1/2$ and $z=1$. For the free theory these exponents remain the
same in all dimensions.

From an experimental point of view this limit can be realized in the
case of merging (coupling through tunneling) two one-dimensional
Bose gases; indeed as it has been pointed out in \cite{nostro} the
SG hamiltonian  (\ref{sgham}) describes the merging process through
the identification of $\beta=\sqrt{2 \pi /K}$, with $K$ the
Luttinger parameter for each Bose gas,  and $\lambda(t)$
proportional to the tunneling strength between the two tubes. In the
limit of weakly interacting gases $K\to \infty$, hence $\beta \to 0$
and the process is described by the  Hamiltonian (\ref{SGqua}).

Let us consider a sudden quench in the Hamiltonian (\ref{SGqua})
where $\lambda$ is changed in time as: $\lambda(t)=\lambda_f
\Theta(t)$, where $\Theta(t)$ is the step function. It is convenient
to switch to the representation in terms of the creation and
annihilation operators $a^\dag$ and $a$, such that $[a,a^\dag]=1$,
in order to work with the number-state vectors:
$|n\rangle=(a^\dag)^n/\sqrt{n!}|0\rangle$. In particular, for a
fixed $\lambda$  those operators are:
\beq
a_q(\lambda)=\frac{1}{(4 \kappa(\lambda))^\frac{1}{4}}\left(\Pi_q+ i\sqrt{\kappa_q(\lambda)}\;\phi_q\right),\\
a^\dag_q(\lambda)=\frac{1}{(4 \kappa(\lambda))^\frac{1}{4}}\left(\Pi_q^\dagger- i\sqrt{\kappa_q(\lambda)}\;\phi_q^\dagger\right),
\eeq
for each mode $q$, such that the Hamiltonian in this basis becomes:
$H=\sum_q \sqrt{\kappa(\lambda)}(a_q^\dag(\lambda) a_q(\lambda) +
1/2)$.  For uniform quenches all $q$-modes are independent from each
other and the dynamics of the system is factorizable. We are
interested in evaluating the effects of the sudden quench on the
population of the excited levels, therefore we want to express the
initial ground state $|0_q\rangle_{0}$ before the quench
($\lambda=0$) in terms of the eigenstates $\{
|n_q\rangle_{\lambda_f}\}$ of the final Hamiltonian
($\lambda=\lambda_f$):
\be
|0_q\rangle_{0}=\sum_n c_{2n}(q) |2n_q\rangle_{\lambda_f},
\ee
where we used the fact that only even modes are excited in the
quench process because of the parity conservation. The coefficients
$c_{2n}(q)$ can be evaluated expressing the operator $a_q(0)$ in
terms of $a_q^\dag(\lambda_f)$ and $a_q(\lambda_f)$:
\beq\label{c2n}
c_{2 n}(q)& = & (-1)^n \sqrt{\frac{(2n-1)!!}{(2 n)!!}}\left(\frac{\sqrt{\kappa_q(\lambda_f)}-\sqrt{\kappa_q(0)}}{\sqrt{\kappa_q(\lambda_f)}+\sqrt{\kappa_q(0)}}\right)^n  \nonumber \\
 & & \times \sqrt{\frac{2 \, (\kappa_q(\lambda_f)\kappa_q(0))^\frac{1}{4}}{\sqrt{\kappa_q(\lambda_f)}+\sqrt{\kappa_q(0)}}}.
\eeq
One can check that these coefficients are properly normalized:
$\sum_n |c_{2 n}(q)|^2=1$.  Physically each state $|2 n_q\rangle$
represents $2 n_q$ quasiparticles with opposite and equal momenta
$q$ and $-q$. Therefore the average number of excitations in each
mode is $n_{\rm ex}(q)=\sum_n 2n |c_{2n}(q)|^2$. We can also
formally define the probability that the mode with momentum $q$ is
excited as:
\be\label{pexq}
p_{\rm ex}(q)=\sum_{n \neq 0} |c_{2 n}(q)|^2.
\ee
The \emph{total probability of exciting the system} $P_{\rm ex}$ is found as the complementary of the joint probability that no mode is excited after the quench:
\be\label{Pex0}
P_{\rm ex}=1-\prod_{q>0} |c_{ 0_q}|^2=1-\prod_{q>0}(1-p_{\rm ex}(q)).
\ee
In the limit $\sum_q p_{\rm ex}(q)\ll 1$ we see that $P_{\rm ex}(q)\approx \sum_q p_{\rm ex}(q)$.

From Eq.~(\ref{c2n}) we find the condition  $p_{\rm ex}(q)\ll 1$ is satisfied if we impose:
 \be
 \frac{\sqrt{\kappa_q(\lambda_f)}-\sqrt{\kappa_q(0)}}
 {\sqrt{\kappa_q(\lambda_f)}+\sqrt{\kappa_q(0)}}=\frac{\sqrt{q^2+2\lambda_f \beta^2}-q}{\sqrt{q^2+2\lambda_f \beta^2}+q} \ll 1
 \ee
that implies $2\lambda_f \beta^2 \ll q^2$. In this limit we have
$p_{\rm ex}(q)\approx |\lambda_f|^2\beta^4/(8q^4)$. For a system
with  size $L$ the lowest nonzero momentum $q$ is $2\pi \over L$ and
therefore the probabilities that $p_{\rm ex}(q)\ll 1$ for all $q$ is
equivalent to:
 \be\label{lambda_small}
 \lambda_f \ll \lambda_s=\frac{1}{2 \beta^2}\left(\frac{2 \pi}{L} \right)^2.
 \ee
It is obvious that this condition also ensures that $\sum_q p_{\rm
ex}(q)\ll 1$. In this limit the total probability for the system to
be excited is:
\be\label{Pexbos}
P_{\rm ex}\approx \sum_q p_{\rm ex}(q)\approx \sum_q \frac{\beta^4\lambda_f^2 }{8 q^4}=\frac{L^4 \beta^4 \lambda_f^2}{8 (2 \pi)^4}\zeta(4),
\ee
where $\zeta(k)$ is the Riemann's Zeta function. It is easy to check
that the coefficient multiplying $L \lambda_f^2$ is precisely the
fidelity susceptibility $\chi_f$ evaluated in the massless limit
$\lambda=0$. The superlinear scaling of the excitation probability
$P_{\rm ex}/L\sim L^3$ is consistent with the one anticipated in
Ref.~\cite{quench_short} (see also Sec.~\ref{sec:gen}):
$\chi_f(0)\sim L^{-d+2/\nu}=L^3$.

For quenches of larger amplitude, when $\lambda_f>\lambda_s$, as we
already discussed in Sec.\ref{sec:gen} the probability of exciting
the system $P_{\rm ex}$ is almost unity. This indicates that $P_{\rm
ex}$ can not differentiate between different excited states. This
comes from the fact that even if we excite a single pair of
quasiparticles, the state becomes immediately orthogonal to the
ground state and $P_{\rm ex}=1$. At the same time physically this
state is almost indistinguishable from the ground state since a
single pair of excited quasiparticles can not affect any
thermodynamic properties of the system. Let us analyze instead that
following object:
\be\label{Spexbos}
\frac{\sum_q p_{\rm ex}(q)}{L}\approx  \frac{1}{2 \pi} \int_0^{\infty} d q\sum_{n_q\neq 0} |c_{2 n(q)}|^2\approx 0.036 \sqrt{\lambda_f \beta^2},
\ee
which represents the sum of probabilities of excitations of
different momentum modes. In general such object can be defined
first evaluating the two-particle density matrix in the momentum
space and then finding probability that the corresponding mode is
excited. To evaluate Eq.~(\ref{Spexbos}) we used the fact that the
sum over momenta can be converted into the integral and for small
amplitude quenches the limits of integration can be extended to
$[0,\infty)$. The scaling dependence in (\ref{Spexbos}) agrees with
$\sim |\lambda_f|^2 \chi_f(\lambda_f)=|\lambda_f|^{d \nu}$ and hence
it illustrates that the fidelity susceptibility actually describes
the behavior of $\sum_q p_{\rm ex}(q)$ rather than $P_{\rm ex}$.
Only for infinitesimally small quenches $|\lambda_f|\ll
|\lambda_s|$, when these two objects coincide, the fidelity
susceptibility also describes the scaling of $P_{\rm ex}$. This
point can be made even more explicit if we consider a quench
starting and ending in the massive limit $\lambda_i>0$ and
$\lambda_f=\lambda_i+\delta\lambda>0$. Then one can check that as
long as $|\delta\lambda|\ll \lambda_i$ we have $\sum_q p_{\rm
ex}(q)\approx L(\delta\lambda)^2\chi_f(\lambda_i)$. At the same time
a similar scaling for $P_{\rm ex}$ is only valid for
$|\delta\lambda|\lesssim 1/\sqrt{L}$ i.e. in the vanishingly small
interval of quench amplitudes in the thermodynamic limit.

The  sum of probabilities in (\ref{Spexbos}) itself is not a
physical observable. A measurable quantity with the same scaling is
the density of created quasiparticles $n_{\rm ex}=1/L\sum_q n_{\rm
ex}(q)$, where $n_{\rm ex}(q)=\langle a_q^\dagger a_q\rangle$. In
the limit $p_{\rm ex}(q)\ll 1$ the sum in (\ref{pexq}) is dominated
by the first term, $n=1$, corresponding  to the creation of one pair
of quasiparticles with opposite momenta. Thus in this limit $n_{\rm
ex}(q)\approx 2p_{\rm ex}(q)$. From this we conclude that for the
small amplitude quenches $\lambda\lesssim \lambda_s$ we have $n_{\rm
ex}\approx 2p_{\rm ex}$, with $p_{\rm ex}=P_{\rm ex}/L$ given by Eq.
(\ref{Pexbos}). As we will see  the same relation is true if we are
dealing with fermionic systems. Physically this relation comes from
the fact that we are  dealing with few-body operators. Thus in the
lowest order of perturbation theory the coupling through the quench
dynamics leads to generating a small number of quasiparticles in
each momentum mode. If the system is noninteracting (or more
generally integrable) then the number of these quasiparticles is
conserved in time after the quench. In nonintegrable systems $n_{\rm
ex}$ changes in time due to scattering processes. Therefore the
scaling of $n_{\rm ex}$ with $\lambda_f$ is expected to be valid
only either at times shorter than the relaxation time, or if the
generated defects are topologically protected like in the
Kibble-Zurek mechanism~\cite{kz1, kz2}. We note that the scaling of
the heat is insensitive to the relaxation processes because the
energy in a closed system is conserved after the quench process. The
same is true about the (diagonal) entropy.

For finite quench amplitudes $\lambda\geq \lambda_s$ the expressions
for $n_{\rm ex}(q)$ and $1/L\sum_q p_{\rm ex}(q)$ are different. In
particular, it is easy to check that
\be
n_{\rm ex}(q)={(\sqrt{\kappa_q(\lambda_f)}-\sqrt{\kappa_q(0)})^2\over 4\sqrt{\kappa_q(\lambda_f)\kappa_q(0)}}.
\ee
For $q\gg 2\lambda_f\beta^2$ we find that $n_{\rm ex}(q)\approx
|\lambda|^2\beta^4/(4q^4)\approx 2p_{\rm ex}(q)$. In the opposite
limit $n_{\rm ex}(q)\approx \sqrt{2\lambda\beta^2}/q$ while $p_{\rm
ex}(q)\approx 1$. The difference between $n_{\rm ex}(q)$ and $p_{\rm
ex}(q)$ happens because the states with more than two quasiparticles
per each mode are excited for small momenta. Such states contribute
differently into the probability of excitation and the quasiparticle
density. For the total density of quasiparticles we thus find:
\be\label{Snexbos}
n_{\rm ex} \approx \frac{1}{2 \pi} \int_\frac{2 \pi}{L}^{\Lambda} d q\, n_{\rm ex}(q)
\approx {\sqrt{2 \lambda_f \beta^2}\over 4\pi}\log\left({\lambda_f \beta^2L^2\over 2\pi^2}\right),
\ee
where $\Lambda$ is the upper momentum cutoff. Note that this
integral gives additional logarithmic dependence on both the quench
amplitude and the system size. This dependence appears because of
the infrared divergence of the density of excited quasiparticles at
small $q$. It is very similar in nature to the onset of the
non-adiabatic regime for slow quenches~\cite{tolianature} and is due
to the bunching effect. As we will argue later this divergence
disappears in dimensions higher than one where the scalings of both
$n_{\rm ex}$ and $\sum_q p_{\rm ex}(q)/L^d$ with the quench amplitude become
identical. In one dimension  the two scalings agree apart from a
weak logarithmic correction.

Similarly we can analyze the heat density, or equivalently the
non-adiabatic energy change in the system due to the quench. For
$\lambda <\lambda_s$  we can find, similarly to Eq.~(\ref{Pexbos}):
\be\label{Qbos}
Q=\frac{1}{L}\sum_{q>0} \sqrt{\kappa_q(\lambda_f)}n_{\rm ex}(q) \approx
\frac{1}{L}\sum_{q>0} \frac{\beta^4\lambda_f^2 }{4 q^3}=\frac{L^2 \beta^4 \lambda_f^2}{4 (2 \pi)^3}\zeta(3).
\ee
In the opposite limit $\lambda >\lambda_s$  we have:
\beq\label{SQbos}
Q & \approx & \frac{1}{2 \pi} \int_\frac{2 \pi}{L}^{\Lambda} d q \;\sqrt{q^2+2 \lambda_f \beta^2} n_{\rm ex}(q)\nonumber\\
& \approx &{\lambda_f \beta^2\over 2\pi}\log\left({\lambda_f \beta^2L^2\over 2\pi^2}\right).
\eeq
Up to the logarithmic correction in the above equation, these
results agree with the prediction in Sec. \ref{sec:gen}: $Q\sim
\lambda^2\chi_E(\lambda)$ with $\chi_E(\lambda)\sim L^{2/\nu-(d+z)}$
for $\lambda<\lambda_s$ and $\chi_E(\lambda)\sim
|\lambda|^{(d+z)\nu-2}$ for $\lambda>\lambda_s$. We point out that
if we consider an opposite process where we start with a small
initial coupling $\lambda_i$ and end at the critical point, then the
expression for $n_{\rm ex}$ will remain the same, while the
expression for the heat for $\lambda>\lambda_s$ will change because
the energy is now evaluated at the critical point. This will remove
the extra logarithmic divergence and we will simply get $Q\propto
\lambda_i\beta^2$.

Finally let us consider the density of the \textit{(d-)entropy}
generated in the quench. Like $P_{\rm ex}$ it gives the measure of
the excitation of the system, however, it is explicitly extensive
and has  well defined scaling properties both for
$\lambda<\lambda_s$ and $\lambda>\lambda_s$. Unlike $n_{\rm ex}$,
the entropy (like the heat) can be defined for any system,
integrable or not. Because all momentum modes are independent we
find that $S_d=1/L\sum_{q>0} s_q$, where
\be
s_q=-\sum_{n\geq 0} |c_{2 n}(q)|^2 \log\left(|c_{2 n}(q)|^2\right).
\ee
For $\lambda\ll \lambda_s$ the entropy of each mode is dominated by
the lowest excitation: $s_q\approx -|c_2(q)|^2\log
|c_2(q)|^2|=-p_{\rm ex}(q)\log p_{\rm ex}(q)$, therefore:
\be
S_d\approx -\frac{L^3 \beta^4 \lambda_f^2}{8 (2 \pi)^4}\zeta(4)\log\left({|\lambda_f|^2\beta^4 L^4\over 8 (2\pi)^4}\right).
\ee
We note that for $\lambda<\lambda_s$ the argument of the logarithm
is smaller then one so the entropy is positive as it should. In the
opposite limit $\lambda>\lambda_s$ we find:
\be
S_d\approx 0.14 \sqrt{\lambda_f\beta^2}.
\ee
So we see that the entropy has the same scaling as $n_{\rm ex}$ and $1/L\sum_q p_{\rm ex}(q)$.

A similar analysis can be done also in the case of a linear quench,
$r=1$ and $\upsilon=\delta$. As it has been shown in
Ref.~\cite{kolkata}, the scalings for the density of excited
quasiparticles: $n_{\rm ex}\sim \delta^{1/3}$  and for the heat,
$Q\sim \delta^{2/3}$, in the case of quenches with rate $\delta\gg
1/L^3$ agree with the predictions in Eqs. (\ref{nex3}) and
(\ref{Qex3}). Similarly it is easy to check  the correct scalings in
the other regime, $\delta\ll 1/L^3$. For instance in this regime
using the expression for the number of excitation per mode $q$ (see
Eq. (103) in Ref.~\cite{kolkata}), we obtain:
\be
n_{\rm ex}= \frac{1}{L}\sum_q n_{q}= \frac{1}{L}\sum_q \frac{(\beta^2\delta)^2 }{16 q^6}=\frac{L^5 (\beta^2\delta)^2 }{16 (2 \pi)^6}\zeta(6),
\ee
confirming the general scaling (\ref{nex03}), which for $\nu=1/2$, $d=1$, $z=1$, and $r=1$ gives $n_{\rm ex}\sim L^5 \delta^2$.

{\em Generalization to higher dimensions.} Let us note that the
Hamiltonian~(\ref{SGqua}) can be analyzed in any spatial dimension.
The quench process simply describes the response of the free bosonic
theory to a sudden turn on of the mass term. Since all the momentum
modes are independent from each other the expressions for
$c_{2n}(q)$, $p_{\rm ex}(q)$, $n_{\rm ex}(q)$ and $s_q$ remain the
same. The only difference with one dimension is that we have to sum
over all modes in the d-dimensional space. This introduces an
additional density of states factor $\rho(q)\propto q^{d-1}$ into
all the expressions. Let us focus only on quenches with
$\lambda>\lambda_s$. It is easy to check that for $1<d<4$ one can
rescale the momentum $q=\sqrt{2\lambda_f \beta^2} \eta$ and set the
limits of integration over $\eta$ to $[0,\infty)$ when analyzing
$1/L\sum_q p_{\rm ex}(q)$, $n_{\rm ex}$, and $S_d$. This follows
from the fact that at large $q$ we have $p_{\rm ex}(q)\sim 1/q^4$ so
the integrals over momenta converge  and the upper cutoff can be
sent to infinity. Likewise in all dimensions larger than one there
are no infrared divergencies in all the integrals so the lower limit
of integration over $q$: $q_{\rm min}\approx 2\pi/L$ can be sent to
zero. Then for all these quantities we get the same desired scaling,
e. g. $n_{\rm ex}\approx C_d (\lambda_f\beta^2)^{d/2}$, where $C_d$
is a number which depends only on the dimensionality. Similarly for
the heat for $d\le 3$ we get $Q\approx \tilde C_d
(\lambda_f\beta^2)^{(d+1)/2}$. Above four dimensions (three
dimensions for the heat) the integrals over momentum become
ultraviolet divergent. So for $d>3$ the leading contribution to heat
will be determined by the large momentum asymptotic of the
transition probability:
\be
Q\approx 2{\lambda_f^2\beta^4\over 8} \int {d^dq\over (2\pi)^d}
{q\over q^4}\approx C(\lambda_f^2\beta^4)|\Lambda|^{d-3},
\label{q_high_d}
\ee
where $\Lambda\sim \pi$ is the high-momentum cutoff and $C$ is a
non-universal constant. A similar expression is valid for $n_{\rm
ex}\approx 1/L\sum_q p_{\rm ex}(q)$ and for the entropy above four
dimensions with the cutoff entering in the power $\Lambda^{d-4}$.
For the entropy there is an additional logarithmic factor
$\log(\lambda_f^2\beta^4/\Lambda^4)$. In three (four) dimensions
heat (density of quasiparticles) acquire an additional logarithmic
dependence on the quench amplitude: $Q\sim
\lambda_f^2\beta^4\log(\Lambda/(\lambda_f\beta^2))$. This divergence
indicates the crossover from quadratic to universal leading
asymptotics. In Sec.~\ref{sec:cpt} we already discussed how similar
logarithmic corrections show up in heat in one dimensional SG model
when $K=1$ for $Q$ ($K=3/2$ for $n_{\rm ex}$).

Using the free bosonic theory it is also easy to verify that for
$3<d<5$ the first subleading asymptotic in the expansion of $Q$ in
powers of $\lambda_f$ is non-analytic and cutoff independent. This
follows from the fact that:
\beq
&&Q= \int {d^dq\over (2\pi)^d}  \sqrt{q^2+2\lambda_f\beta^2} \left(n_{\rm ex}(q)-{\lambda_f^2\beta^4\over 4q^4}\right)\nonumber\\
&&+\int {d^dq\over (2\pi)^d}  q {\lambda_f^2\beta^4\over 4q^4}\approx C\lambda_f^2\beta^4|\Lambda|^{d-3}+B |\lambda_f\beta^2|^{d+1\over 2}.\phantom{XX}
\label{q_d}
\eeq
Since $n_{\rm ex}(q)-\lambda_f^2\beta^4/ 4q^4\sim
-\lambda_f^3\beta^6/2 q^6$  the first integral in the above equation
is ultraviolet convergent below five dimensions and we can use the
rescaling $q=\sqrt{2\lambda_f\beta^2} \eta$ and send limits of
integration over $\eta$ to $[0,\infty)$. We can continue and get the
asymptotical expression for the heat in high dimensions similar to
Eq.~(\ref{qexp})

\be
Q=\sum_{2<n<[(d+1)/2]} C_n |\lambda_f|^{n} + B |\lambda_f|^{(d+1)/2}+o(|\lambda|^{(d+1)/2}).
\label{qexp_d}
\ee
We note that this analytic expansion goes in both even and odd
powers of $\lambda_f$ (unlike Eq.~(\ref{qexp})) where only even
powers of $\lambda$ appear. This has to do with the fact that for
the massive bosonic theory (unlike the SG model)
$\lambda\to-\lambda$ is not a symmetry of the Hamiltonian. Moreover,
the theory is well defined only for $\lambda>0$. As in
Eq.~(\ref{qexp}) we find logarithmic corrections when nonanalytic
and analytic powers coincide, i.e. when $(d+1)/2$ becomes an
integer. A very similar expression holds for $n_{\rm ex}$ with the
only difference that in Eq.~(\ref{qexp_d}) one needs to replace
$d+1$ with $d$. We point out that for small amplitude quenches
$\lambda<\lambda_s$ the expansion (\ref{qexp_d}) will remain valid
with the first non-universal analytic terms being unaffected, while
in the last non-analytic terms one needs to substitute
$|\lambda_f|^{(d+1)/2}$ to $\lambda_f^2 L^{3-d}$. Therefore above
three dimensions for quenches of small amplitude the non-analytic
correction to the heat becomes sub-extensive. The same is true for
the density of quasiparticles  where the non-analytic term scales as
$\lambda_f^2 L^{4-d}$.

\subsection{Free massive fermions}
\label{sec:exact:fermi}

Another important application of the SG Hamiltonian~(\ref{sgham}) to
physical systems  feasible in cold atoms experiments, is given by
the description of the loading process of a one-dimensional Bose gas
into a commensurate optical lattice. In this case $\beta$ is related
to the Luttinger parameter through: $\beta=2\sqrt{\pi K}$, and the
amplitude  of the optical lattice  $V(t)$ is directly proportional
to $\lambda$ \cite{nostro,buchler}. This process can be studied
exactly in the Tonks-Girardeau regime, i.e. when $K=1$. In this case
the repulsive interaction between the bosons is infinitely strong
and the particles behave as impenetrable spheres (hard-core bosons).
It is well known that in this limit the system can be mapped into an
equivalent system of free spinless fermions \cite{girardeaus}.
Therefore the dynamical problem  of loading hard-core bosons into a
commensurate optical lattice  can instead be approached with the
much simpler analysis of free massive fermions in a periodic
potential. We note that this fermionic limit also describes the
transverse field Ising model via the Jordan-Wigner
transformation~\cite{Sachdev_book}. Then the quench dynamics is
performed by suddenly change the transverse field starting at the
critical point.

To understand the dynamics in the TG-gas limit, we need to solve the
Schr\"odinger equation describing free fermions in a periodic
commensurate potential with time dependent amplitude $V(x,t)=V(t)
\cos(2q_f x)$, where $q_f=\pi/a$ is the Fermi momentum and $a$ is
the lattice spacing, which is also equal to the inverse particle
density. In the TG-limit  the sound velocity $v_s=v_f/K$ is equal to
the Fermi velocity $v_f$.
  It is convenient to work in units such that $a=1$, $v_f=1$ and $\hbar=1$.
The free fermion Hamiltonian becomes equivalent to the SG
Hamiltonian~(\ref{sgham}) if we additionally restrict the analysis
to the two lowest bands of the Brillouin zone and linearize the
spectrum close to the Fermi momentum. Then for each momentum the
loading process is described by a Landau-Zener
Hamiltonian~\cite{landauzener} (see also Ref.~\cite{dziarmaga} for
the discussion of the equivalent transverse field Ising model):
\be\label{hLZ}
\HH_k\approx \left[\begin{matrix} V(t)/ 2 & k \\ k & -V(t)/2 \end{matrix}\right],
\ee
where we defined the momentum $k= q-q_f$, so that  is measured from
the Fermi momentum. By matching this spectrum with the one of the SG
model we find that $\lambda=V/(4\pi)$. The extra factor of $4 \pi$
appears because of the details of the bosonization
procedure~\cite{mora,cazalilla}.

The  Hamiltonian (\ref{hLZ}) can be diagonalized on the basis of its
eigenvectors: $|+_k\rangle$ and $|-_k \rangle$, with respectively
positive and negative energy. To study the probability of exciting
the system due to a sudden quench of the coupling $V$:
$V(t)=V_f\theta(t)$, we need to evaluate the overlap of the initial
ground state  $|-_k\rangle_0$ with the final excited state
$|+_k\rangle_{V_f}$.  Therefore $p_{\rm ex}(k)=|_{V_f}\langle
+_k|-_k\rangle_0|^2$. In particular we find:
\be\label{pTG}
p_{\rm ex}(k)=\frac{1}{2}\frac{V_f^2+k^2 -V_f |k|+(V_f-|k|)\sqrt{V_f^2+k^2}}{V_f^2+k^2+V_f\sqrt{V_f^2+k^2}}.
\ee
This transition probability has the following asymptotics: $p_{\rm
ex}(k)\approx V_f^2/(4k^2)$ for $k\gg V_f$ and $p_{\rm ex}(k)\approx
1/2$ for $k\ll V_f$. As in the previous case of free bosons, the
elementary excitation correspond to the creation of particle-hole
pairs moving with opposite momenta (recall that annihilating a
particle in the lower band can be viewed as creating a hole there
with an opposite momentum). However, unlike with bosons, we can not
create more than one pair for each momentum state due to the Pauli
exclusion principle. This means that $1/L\sum_k p_{\rm ex}(k)$ and
$n_{\rm ex}$ are always identical up to a factor of two. As in the
bosonic case the crossover scale $V_s$ can be defined from the
requirement that $\sum_k p_{\rm ex}(k)\ll 1$, which gives $V_s\sim
4\pi/L$ (or equivalently $\lambda_s\sim 1/L$). For $V<V_s$ the total
probability to excite the system is small so that $P_{\rm ex} \simeq
\sum_k p_{\rm ex}(k)$. This characteristic scale agrees with the
expected one $\lambda_s \simeq L^{-1/\nu}$ from the analysis of the
system close to the quantum critical point, since in this case
$\nu=1$. Using the explicit asymptotics for the excitation
probability $p_{\rm ex}(k)$ at small amplitudes $V_f$ we find that
in this case:
\be
{P_{\rm ex}\over L}\approx {n_{\rm ex}\over 2}\approx {V_f^2 L\over 48}={\pi^2\over 3}\lambda_f^2 L.
\ee
For large quenches $V_f\gg V_s$ we find that:
\be
n_{\rm ex}\approx 4 \int_{-\pi}^0 {dk\over 2\pi}\; p_{\rm ex}(k)=\frac{V_f}{\pi}=4\lambda_f.
\ee
We see that as in the case of bosons $n_{\rm ex}$ has the correct
scaling $n_{\rm ex}\sim |\lambda_f|^{d\nu}$ for $\lambda>\lambda_s$
and $n_{\rm ex}\sim \lambda_f^2 L^{2/\nu-d}$ for $\lambda<\lambda_s$
in agreement with Eqs.
 (\ref{nex03}) and (\ref{nex3})  respectively.

Likewise one can evaluate the heat density $Q=2/L\sum_k p_{\rm ex}(k) \sqrt{k^2+V_f^2/4}$. For $\lambda<\lambda_s$ we find:
\beq
Q&=&\frac{2}{L}\sum_k \sqrt{k^2+V_f^2/4} p_{\rm ex}(k)\nonumber\\
&\approx& {V_f^2\over 2\pi}\log(L/2)=8\pi\lambda_f^2\log(L/2),
\label{heat_ferm1}
\eeq
and in the opposite limit $\lambda>\lambda_s$:
\be
Q={V_f^2\over 8\pi}\log(2\pi/V_f)=-2\pi\lambda_f^2\log(2\lambda_f).
\label{heat_ferm2}
\ee
This scaling of heat agrees with the general expectations of Eqs.
(\ref{Qex03}) and (\ref{Qex3}) and the results obtained in the
previous sections using the adiabatic perturbation theory. Since we
have $(d+z)\nu=2$, there is an additional logarithmic correction
indicating the crossover from quadratic to nonanalytic scaling in
the leading asymptotic of the scaling of $Q$ with $\lambda_f$. Note
that the difference in scaling between $\lambda<\lambda_s$ and
$\lambda>\lambda_s$ only appears in the logarithmic term.

Finally we can evaluate the entropy density. Let us explicitly quote
only the result for $\lambda>\lambda_s$:
\beq
&&S_d=-\frac{1}{L}\sum_k \bigl[[p_{\rm ex}(k) \log(p_{\rm ex}(k))\nonumber\\
&&~~~+(1-p_{\rm ex}(k)\log(1-p_{\rm ex}(k))\bigr]={V_f\over 2}=2\pi\lambda_f.\phantom{XX}
\eeq
The scaling of the entropy is thus again in accord with the exact
solution. It is easy to check that for $\lambda<\lambda_s$ the
entropy density scales as $S\sim \lambda_f^2 L\log (\lambda_f L)$,
i.e. it is super-extensive.

As in the bosonic case, these predictions can be also checked in the
case of slow quenches. In Ref.~\cite{kolkata} it has been shown that
for the linear quench $\lambda(t)=\delta t$ one finds $n_{\rm
ex}\sim \sqrt{\delta}$ for $\delta \gg 1/L^2$, in agreement with Eq.
(\ref{nex3}), while for $\delta \ll 1/L^2$ one can find that (see
Eq. (108) and Appendix A in \cite{kolkata} for the expression for
$n_q$):
\be
n_{\rm ex}= \frac{1}{L}\sum_k n_k = \frac{1}{L}\sum_k \frac{4 \delta^2 }{k^4}=\frac{4 L^3  \delta^2 }{ (2 \pi)^4}\zeta(4),
\ee
which confirms the scaling  $n_{\rm ex}\sim L^3 \delta^2$ given by Eq. (\ref{nex03}).

{\em Generalization to higher dimensions.} Like in the case of
bosons one can extend the analysis to higher dimensional free
fermionic models. For example, the two-dimensional generalization of
the free fermionic theory describes commensurate fermions in the
honeycomb lattice, which are realized in
graphene~\cite{antonio_rmp}. Such lattices can be also in principle
realized in cold atoms using multiple laser beams~\cite{solomon}.
The  quench process corresponds to the distortion of the lattice,
which opens a gap in the spectrum. Quench dynamics in similar setup
was recently analyzed in Ref.~\cite{dutta2009b,balazs}.

Unlike in the bosonic limit, where $\nu=1/2$, here we have $\nu=1$.
This implies that we expect that the quadratic scaling for the
quasiparticle density and entropy emerges above two dimensions and
for the heat above one dimension. This is indeed the case since the
excitation probability at large momenta scales as $p_{\rm ex}(k)\sim
1/k^2$ (versus $1/k^4$ in the case of bosons). To understand the
general structure of the heat density above one dimension we can
more closely examine the expansion of the transition probability
$p_{\rm ex}(k)$ at large $k$. In particular, from Eq.~(\ref{pTG}) we
find:
\be
p_{\rm ex}(k)\approx {V_f^2\over 4k^2}-{3V_f^4\over 16 k^4}+{5V_f^6\over 32 k^6}+\dots.
\ee
As we see the expansion is in  even powers of $V_f$ which reflects
the symmetry of the problem with respect to $V_f\to-V_f$. For
$1<d<3$ we can rewrite the  heat as
\beq
&&Q\approx 2 M\int_{-\pi}^0 {d^dk\over (2 \pi)^d} \sqrt{k^2+V_f^2/4} \left(p_{\rm ex}(k)-{V_f^2\over 4k^2} \right)\nonumber\\
&&~~~+ 2 M \int_{-\pi}^0 {d^dk \over (2 \pi)^d } {V_f^2\over 4k},
\eeq
where $M$ is the number of independent light cones in the system
($M=2$ for graphene). The second contribution gives a non-universal
quadratic term, while in the first integral we can rescale the
variables $k\to V_f\eta$ and send the limits of integration over
$\eta$ to $[0,\infty)$. This contribution thus gives the
non-analytic term $|V_f|^{(d+z)\nu}$. A similar analysis can be
extended to higher dimensions and as a result we get the following
expansion:
\be
Q=\sum_{1<n<[(d+1)/2]} C_n |\lambda_f|^{2n} + B |\lambda_f|^{d+1}+o(|\lambda|^{d+1}).
\label{qexp_df}
\ee
The coefficients $C_n$ are non-universal (cutoff dependent) while
the coefficient $B$ is universal and can be obtained from the low
energy description of the system. Similar considerations can be
applied to the density of quasiparticles resulting in the
expression similar to Eq.~(\ref{qexp_df}) with $d+1\to d$. For small
amplitude quenches $\lambda<\lambda_s$ only the universal term in
the expansion above gets modified with $|\lambda_f|^{d+1}\to
\lambda_f^2 L^{1-d}$. As we already mentioned  in the points where
the power of the non-analytic term crosses the one of the analytic
contribution ($d+1=2n$) there are additional logarithmic corrections
to the heat.

\subsection{Finite temperature quenches}
\label{sec:temp}

Let us extend the analysis of these thermodynamic quantities to the
case of a \textit{finite initial temperatures}. We suppose to
perform the same quench process but now preparing the system
initially in a thermal state at temperature $T$. We will assume that
the system is isolated during the dynamical process. For
instantaneous quenches this assumption is generally always
satisfied. For slow quenches this assumption is only valid if we are
interested in the time scales shorter than the relaxation time with
the thermal bath. This is the usual requirement in order to be in
the adiabatic limit in conventional thermodynamics~\cite{LL5} and it
is usually well satisfied in cold atoms. We note that other studies
have addressed the case of systems where the contact with a thermal
bath during the dynamical process is essential~\cite{patane}. The
advantage of focusing on the setup where the temperature enters only
through the initial conditions and does not affect the equations of
motion is that one can expect that the scaling of various quantities
will remain universal. We note that finite temperature dynamics in
this setup were studied earlier in Refs.~\cite{tolianature, kolkata}
for slow linear quenches and in Ref.~\cite{cardy_temp} for sudden
quenches.

\subsubsection{Free massive bosons.}

For the case in consideration of a system of independent harmonic
oscillators (see Eq.~(\ref{SGqua})), a finite initial temperature
leads to the occupation of the eigen-modes according to the
Bose-Einstein distribution:
\be
n_T(q)={1\over \exp[\beta\sqrt{\kappa_q(0)}]-1}={1\over \exp[\beta q]-1}.
\ee
This initial occupation enhances the transition probabilities. In
Ref.~\cite{tolianature} it was shown that for an arbitrary time
dependence of $\lambda(t)$, the mode occupation with momentum $q$ at
finite temperature is related to the one at zero temperature via:
\beq
n_{\rm ex}^{\rm tot}(q)&=&n_{\rm ex}(q)\coth\left[{\sqrt{\kappa_q(0)}\over 2T}\right]+n_T(q)\nonumber\\
&=&n_{\rm ex}(q)\coth\left[{|q|\over 2T}\right]+n_T(q).
\label{bosCoth}
\eeq
where $n_{\rm ex}^{\rm tot}(q)$ is the total number of excited
quasiparticles during the dynamical process, that  is indeed equal
to the initial thermal population of this mode ($n_T(q)$), plus the
number of particles generated in the identical process at zero
temperature multiplied by a hyperbolic cotangent factor. Therefore
we can define the excitations $n_{\rm ex}^{T}(q)$ created only by
the dynamical process at a finite temperature as the difference
$n_{\rm ex}^{\rm tot}(q)-n_T(q)$. It is immediately clear that the
hyperbolic factor gives an enhancement of the transitions since at
$T\gg q$ it scales as $2T/q\gg 1$. The result~(\ref{bosCoth}) can be
seen either from the thermal average of the energy of each mode $q$
as it was recently pointed out in Ref.~\cite{cardy_temp}, or writing
the initial thermal state in the Wigner form, as shown in
Ref.~\cite{tolianature}.

Let us now analyze how this additional factor affects the scaling of
the quantities of interest. To simplify the discussion we will focus
only on the heat density. We will also assume that the temperature
is large compared to both the energy associated with the quench
amplitude $T\gg \sqrt{\lambda_f \beta^2}$ and the finite size
quantization $T\gg 1/L$. Otherwise we will return to the zero
temperature asymptotics.

For $\lambda<\lambda_s$ by simple modification of Eq.~(\ref{Qbos}) we find

\be\label{QbosT}
Q\approx \frac{2T}{L }\sum_{q>0} \frac{\beta^4\lambda_f^2 }{4 q^4}=\frac{T L^3 \beta^4 \lambda_f^2}{2 (2 \pi)^4}\zeta(4).
\ee
In the opposite limit $\lambda >\lambda_s$  instead of Eq.~(\ref{SQbos}) we find:
\be\label{SQbosT}
Q \approx \frac{2T}{L }\sum_{q>0} {2\lambda_f\beta^2\over q^2}\approx {1\over 6} TL\lambda_f\beta^2.
\ee
For small quenches the result (\ref{QbosT}) is fully consistent with
the scaling prediction of the Eq. (\ref{Qex03}): $Q\sim \lambda_f^2
L^{2/\nu-d}$. While for larger quench amplitudes we get a deviation.
Instead of $Q\sim |\lambda_f|^{d\nu}=\sqrt{|\lambda_f|}$ we get the
scaling $Q\sim |\lambda_f|L$. The origin of this discrepancy is the
infrared divergence of the sum in Eq.~(\ref{SQbosT}) at small
momenta. This divergence is similar to the one appearing in linear
quenches, which leads to the non-adiabatic
regime~\cite{tolianature}. As it is easy to check this divergence
disappears in higher dimensions and the predicted general scaling is
restored. Indeed for $2<d<4$ and for $\lambda>\lambda_s$ one can
substitute the summation over momenta in the general expression for
$q$ by an integration, change the variables
$q=\sqrt{2\lambda_f\beta^2}\eta$ and extend the limits of
integration over $\eta$ to $[0,\infty)$. This immediately implies
that $Q\sim T|\lambda_f\beta^2|^{d/2}$. Above four dimensions the
integral over momenta becomes ultra-violet divergent. In this case
the transitions are dominated by high momenta for which the initial
thermal occupation is not important (if the temperature is much
smaller than the high-energy cutoff: $T\ll\pi$ in our units) and we
are back to the zero temperature non-universal
result~(\ref{q_high_d}). It is easy to check that in this case (as
at zero temperatures) the universal nonanalytic correction to $Q$
becomes subleading thus instead of Eq.~(\ref{qexp_d}) we find:
\be
Q=\sum_{2<n<[d/2]} C_n |\lambda_f|^{n} + B T |\lambda_f|^{d/2}+o(|\lambda|^{d/2}).
\label{qexp_d_T}
\ee

To summarize the finite temperature results for the bosonic systems
we see that the substitution $d\to d-z$ always predicts the correct
finite temperature scaling of the heat for $\lambda<\lambda_s$. In
the opposite limit $\lambda>\lambda_s$ this substitution $d\to d-z$
also works in two dimensions and above but fails in one dimension.
The same applies to the density of generated quasiparticles.

\subsubsection{Free massive fermions.}

It is also straightforward to obtain the finite temperature
asymptotics in the fermionic case. Since in this case we are dealing
with a sum of independent two-level systems, at finite temperature
each upper $|+_k\rangle$  and lower $|-_k\rangle$ level is occupied
according to the Fermi distribution:
\begin{displaymath}
f_k^{\pm }=\left(\exp\left[\pm \frac{|k|}{T}\right]+1\right)^{-1},
\end{displaymath}
where we used the convention that the Fermi velocity is one so that
the energy of the two states $\epsilon_{\pm}(k)\approx \pm k$. It is
well known that for the two-level system the transition probability
satisfies the detailed balance, i.e. the transition probability from
the top to the bottom level is the same as for the opposite process
(for systems with more than two levels the detailed balance is
generally not true, see e.g. Ref.~\cite{ap_heat}). Therefore the
number of additionally excited particles in the dynamical process
gets corrected as:
\be
n^T_{\rm ex}(k) = n_{\rm ex}(k) (f_k^{-}-f_k^{+}) = n_{\rm
ex}(k)\tanh\left(\frac{ |k|}{2 T}\right).
\ee
This expression is similar to the one  for the bosonic case
(\ref{bosCoth}), with now the hyperbolic tangent factor replacing
the cotangent one. Contrary to the bosonic case therefore, the
effect of the finite temperatures is to suppress the transitions
since $|\tanh(x)| <1$. At high temperatures $T\gg k$ the suppression
factor is approximately $|k|/(2T)$.

As in the case of bosons let us focus on the scaling of heat. Note
that the main contribution to Eqs.~(\ref{heat_ferm1}) and
(\ref{heat_ferm2}) even at zero temperature comes from high energies
$k\sim\Lambda$. Therefore in the leading order in $\lambda_f$ the
temperature will not significantly modify the expression for the
heat except changing the cutoff in the logarithm:
\be
Q\approx 8\pi\lambda_f^2 \log(\pi/T).
\ee
This scaling is valid both for $\lambda<\lambda_s$ and
$\lambda>\lambda_s$. This is indeed an anticipated result because
the universal power $(d+2z)\nu=3$ is bigger than two. The next
subleading correction in $\lambda_f$ remains universal though and
scales as $\lambda_f^2/L$ for $\lambda<\lambda_s$ and
$|\lambda_f|^3/T$ for $\lambda>\lambda_s$. It is straightforward to
check that this general structure of the expansion extends to higher
dimensions so that for $\lambda>\lambda_s$ instead of
Eq.~(\ref{qexp_df}) we find:
\be
Q=\sum_{1<n<[(d+2)/2]} C_n |\lambda_f|^{2n} + \tilde B |\lambda_f|^{d+2}+o(|\lambda|^{d+2}).
\ee
The non-universal coefficients $C_n$ are temperature independent
(for $T\ll \Lambda$) while the universal coefficient $\tilde B$ is
different from the zero-temperature one. We thus see that the finite
temperature effects result in changing $d\to d+z$ in the universal
part of the expression for heat. It is easy to check that the same
is true for the density of excited quasiparticles.

\section{Quantum geometric tensors and fidelity susceptibility. Mazur inequalities.}
\label{sec:geom_tens}

\subsection{Quantum geometric tensors: generalizations}
In the discussion above we only
considered situations where $\lambda$ was a single component
coupling. In principle one can extend the analysis to more
general setups where ${\bf \lambda}$ is a $M$-component vector in
the parametric space. Within the adiabatic perturbation theory it is
easy to generalize the expression for the transition probability to
the many-body state $|n\rangle$ due to a sudden quench in the
multicomponent case:
\be
|\alpha_n(\lambda)|^2\approx \left|\int d{\bm \lambda'}
{\langle n|\partial_{\bm \lambda'}H|0\rangle\over E_n({\bm \lambda'})-E_0({\bm \lambda'})}\right|^2,
\label{alphan}
\ee
where the contour integral is taken over an arbitrary path
connecting the initial and final couplings. For sudden quenches the
precise path is clearly unimportant. Next we will use the
Cauchy-Schwarz inequality,  which for arbitrary square integrable
functions $f(x),\,g(x)$ states that
\beq
\left|\int f(x) g(x) dx \right|^{2}\leq\int|f(x)|^{2}dx\cdot \int|g(x)|^{2}dx.
\eeq
Choosing $g(x)=1$ and $f(x)$ to be the integrand in Eq.~(\ref{alphan}) we find:
\be
P_{\rm ex}=\sum_{n\neq 0}|\alpha_n(\lambda)|^2\leq M\sum_{\alpha=1}^{M}
\lambda_\alpha\int_0^{\lambda_\alpha} d\lambda'_\alpha Q_{\alpha,\alpha}(\lambda'),
\ee
where $Q_{\alpha,\alpha}(\lambda)$ are the diagonal components of the
quantum geometric tensor $Q_{\alpha,\beta}(\lambda)$ defined in
Ref.~\cite{venuti}. This tensor and its rescaled counterpart
($q_{\alpha\beta}=L^{-d}Q_{\alpha\beta}$) has definite scaling
properties  close to the quantum critical point:
\beq
q_{\alpha\beta}(\lambda)\sim |\lambda|^{\nu\Delta_{Q_{\alpha,\beta}}}
\eeq
where, according to Ref.~\cite{venuti}:
\be
\Delta^{Q}_{\alpha\beta}:=\Delta_{\alpha}+\Delta_{\beta}-2z-d.
\label{delta}
\ee
We remind that $\nu$ is the correlation length critical exponent,
which can in principle depend on the direction of the quench in
the ${\bf \lambda}$ space, and $z$ is the dynamical critical
exponent. For a single-component, $\Delta_\alpha$ is the
scaling dimension of the operator $V=\partial_\lambda
H$. Assuming that $V$ is non-zero in the limit $\lambda\to 0$ from
the scale invariance of the action we find that
$\Delta_\alpha=d+z-1/\nu$, which using Eq.~(\ref{delta}) reproduces
the correct scaling of the probability to excite the system $P_{\rm ex}\sim
|\lambda|^{2}L^{2/\nu}$. Note that the diagonal components of the
geometric tensor give the fidelity susceptibilities $\chi_{f}$ discussed earlier (see also Ref.~\cite{gu_review} for
the review). These observations suggest a close relation between the scalings of $P_{\rm
ex}$ with the quench amplitude, the quantum geometric tensors and
the fidelity susceptibility characterizing the static ground state
properties of the system. A similar analysis shows the connection
between the heat and the energy susceptibility $\chi_E$ (see
Eq.~(\ref{chie})).

The importance of the geometric tensor becomes clear in the contest of
multi-parametric dynamics, when several parameters involved in the
model can depend on time. In this case the integrals in the
expressions for the physical quantities should be replaced by the
line integrals in the parameter space.  One can define a
``response'' in terms of one parameter by driving another parameter. The
corresponding susceptibility of this response is described by
the real parts of the non-diagonal matrix elements of the quantum
geometric tensor. Moreover the imaginary part of the geometric
tensor is responsible for the Berry phase which may appear as a
result of adiabatic evolution in this multi-parameter space. We are
going to come back to these questions in the future.

\subsection{Integrable perturbations and ergodicity}
It is interesting to note that the Cauchy-Schwarz inequality, although it gives
simply an upper bound, it appears to be exact for the scaling functions we
derived  here. On the other hand for some type
of perturbations away from the QCP's (the so-called integrable
perturbations) it seems possible to formulate a {\it lower} bound as
well. It is known ~\cite{mazur} that
for the models which have integrals of motion, the time averages of the
correlation functions have a lower bound and therefore their
dynamics can be {\it non-ergodic}. In particular, if the quantities
$A_{i}$($i=1,\cdots m$) are the integrals of motion, $[H,
A_{i}]=0$ $ \forall i$, defined such that $\langle A_{i}\rangle =0$,
then the dynamical correlation function of the operator $X(t)$ is
bounded from below as:
\be
\lim_{T\rightarrow\infty}\frac{1}{T}\int_{0}^{T}\langle X(t)
X(0)\rangle dt\geq \langle X{\bf A}\rangle\cdot\langle{\bf A}{\bf
A}\rangle^{-1}\cdot\langle X{\bf A}\rangle
\ee
where $\langle{\bf A}{\bf A}\rangle$ is the $m\times m$ matrix with
elements $\langle A_{i} A_{j}\rangle$. Provided that this matrix can be
diagonalized by some unitary transformation and that the Hamiltonian is
identified with one of the operators $A_{i}$, say $H\equiv A_{1}$, we
obtain $\langle \tilde{A}_{i}\tilde{A}_{j}\rangle/\langle
\tilde{A}_{i}^{2}\rangle=\delta_{ij}$, where the notation
$\tilde{A}$ stands for the transformed operator, and therefore
~\cite{mazur}
\be
\lim_{T\rightarrow\infty}\frac{1}{T}\int_{0}^{T}\langle X(t)
X(0)\rangle dt\geq\sum_{k=1}^{m}\frac{\langle
X\tilde{A}_{k}\rangle^{2}}{\langle\tilde{A}_{k}^{2}\rangle}\geq\frac{\langle
X H\rangle^{2}}{\langle H^{2}\rangle}.
\label{mazur}
\ee
We implicitly assume that our Hamiltonian is defined in such a way
that $\langle H\rangle=0$.

The sine-Gordon model represents an integrable deformation
(perturbation) of the conformal field theory. Therefore it has many
(infinite number) integrals of motion. Taking $X(t)\equiv
\int_{0}^{L}\cos(\beta\phi(x,t))$ we obtain a lower bound for the
{\it time average} of the energy susceptibility
\beq
\chi_{E}^{(t)}=\lim_{T\rightarrow\infty}\frac{1}{T}\int_{0}^{T}\int_{0}^{L}dx\,
G(x,\tau) d\tau.
\eeq
By the Mazur inequality (\ref{mazur}) this susceptibility has a
lower bound, which is nonzero. I.e. $\chi_E$ can either be finite or
diverge. The Mazur inequalities above can be also directly
generalized for other time averages of the susceptibilities like
$\chi_{f}^{(t)}$ defined in Eq.~(\ref{chi_f}) as well as higher
order adiabatic susceptibilities ~\endnote{A counter-example where
$\chi_f(0)$ vanishes is the Richardson model (private communications
with B. Gut, D. Baeriswyl, and R. Barankov). However, this model is
effectively classical with suppressed quantum fluctuations. The full BCS
model has positive $\chi_f$.}.

\section{Conclusions}

In this work we analyzed the quench dynamics starting (ending) at the quantum
critical point focusing on the sine-Gordon model. We derived the universal
scalings of such quantities as the probability of exciting the system ($P_{\rm
ex}$), density of excited quasiparticles ($n_{\rm ex}$), entropy and heat
densities ($S_d$ and $Q$) with the quench amplitude. These scalings are
fully determined by the critical exponents $z$ and $\nu$ characterizing
the quantum critical point and agree with general expectation presented in
\cite{quench_short}.

In particular for the type of quenches where the tuning
parameter changes as a power law near the quantum critical point:
$\lambda(t)\approx \upsilon t^r/r!$,  we showed that  the scalings of
$P_{\rm ex}$, $n_{\rm ex}$, and $S_d$ are associated with the singularities of
generalized adiabatic susceptibilities $\chi_{2r+2}(\lambda)$
of order $2r+2$ (see Eq.~(\ref{chim})), while if the quench ends at the
critical point the scaling of $Q$ is associated with the singularity of
$\chi_{2r+1}$. We note that for $r>0$ it is sufficient to have gapless
systems, not necessarily quantum critical point, in order to observe these
singularities.

For the two limits where the elementary excitations in the system are
characterized by free bosons or fermions (for the sine-Gordon model these
limits correspond to a particular choice of $\beta$, see
Eq.~(\ref{sgham})) we generalized our results for the finite temperature
quenches. We showed that the structure of the singularities remains the
same except that in the scaling relations for the density of
quasiparticles and heat one needs to substitute the dimensionality $d\to d-z$
for bosons (simultaneously multiplying $Q,\; n_{\rm ex}$ by the
temperature) and for $d\to d+z$ for fermions (simultaneously dividing
$Q,\; n_{\rm ex}$ by the temperature). This changes are a direct
manifestation of the  statistics of the quasiparticles: bunching of bosons enhancing
non-adiabatic effects and anti-bunching of fermions suppressing the
transitions.

We believe our results can be directly tested both numerically and
experimentally, especially in setups realizable with cold atoms.

\subsection*{Acknowledgements}
We acknowledge discussion with D.~Baeriswyl, R.~Barankov and B.
~Gut. We also thank R.~Barankov for pointing out the correct
normalization \cite{Mussardo-book} for the rapidities variables in
Eq.(\ref{matr_el}). The work was supported by  AFOSR YIP, NSF: DMR
0907039, and Sloan Foundation. C.~D.~G. acknowledges the support of
I2CAM: DMR-0645461. V.G. was supported by the Swiss National Science
Foundation.

\bibliography{quenchSG}

\end{document}